\documentclass[structabstract]{aa}
\usepackage{graphicx}
\usepackage{txfonts}
\usepackage{natbib}
\bibpunct{(}{)}{;}{a}{}{,} 

\def\N2H{N$_2$H$^+$}

\def\s{$\pm$}

\def\s12{\mbox{$S_{\rm 1.2mm}$}}

\newcommand{\beq}{\begin{equation}}
\newcommand{\eeq}{\end{equation}}
\newcommand{\bdi}{\begin{displaymath}}
\newcommand{\edi}{\end{displaymath}}

\begin{document}


\title{ On the shape of the mass-function of dense clumps in the Hi-GAL fields }

\subtitle{II. Using Bayesian inference to study the clump mass function }

\author{
L. Olmi \inst{\ref{inst1},\ref{inst2}} \and
D. Angl{\'e}s-Alc{\'a}zar\inst{3} \and
D. Elia\inst{4} \and
S. Molinari\inst{4} \and
M. Pestalozzi\inst{4} \and
S. Pezzuto\inst{4} \and
E. Schisano\inst{\ref{inst5},\ref{inst4}} \and
L. Testi\inst{\ref{inst6},\ref{inst1}} \and
M. Thompson\inst{7}
}

\institute{
	  INAF, Osservatorio Astrofisico di Arcetri, Largo E. Fermi 5,
          I-50125 Firenze, Italy,  \email{olmi.luca@gmail.com} \label{inst1}  
\and
	  University of Puerto Rico, Rio Piedras Campus, Physics Dept., Box 23343, 
	  UPR station, San Juan, Puerto Rico, USA  \label{inst2} 
\and
          Department of Physics, University of Arizona, 1118 E. 4th Street, Tucson,
          AZ 85721, USA \label{inst3}
\and
	  Istituto di Fisica dello Spazio Interplanetario - INAF, via Fosso del Cavaliere 100, 
	  I-00133 Roma, Italy \label{inst4}
\and
	  Infrared Processing and Analysis Center, California Institute of Technology, Pasadena, 
	  CA 91125, USA \label{inst5} 
\and
          ESO, Karl-Schwarzschild-Str. 2, 85748 Garching bei München, Germany \label{inst6}
\and
	  Centre for Astrophysics Research, University of Hertfordshire, College Lane, 
	  Hatfield, AL10 9AB, UK \label{inst7} 
          }

\date{Received; accepted }


\abstract
{Stars form in dense, dusty clumps of molecular clouds, but little is known
about their origin, their evolution and their detailed physical properties.
In particular, the relationship between the mass distribution of these clumps
(also known as the ``clump mass function'', or CMF)
and the stellar initial mass function (IMF), is still poorly understood.
}
{In order to better understand how the CMF evolve toward 
the IMF, and to discern the ``true'' shape of the CMF, 
large samples of  bona-fide pre- and proto-stellar clumps are required.
Two such datasets obtained from the Herschel infrared GALactic Plane 
Survey (Hi-GAL) have been described in paper I.
Robust statistical methods are needed in order to infer the
parameters describing the models used to fit the CMF, and to 
compare the competing models themselves.
}
{In this paper we apply Bayesian inference to the analysis of the CMF of the two 
regions discussed in Paper I. First, we determine the Bayesian posterior probability 
distribution for each of the fitted parameters. Then, we carry out a quantitative 
comparison of the models used to fit the CMF.
}
{We have compared the results from several methods implementing Bayesian inference,
and we have also analyzed the impact of the choice of priors and the influence of various 
constraints on the statistical conclusions for the preferred values of the parameters.
We find that both parameter estimation and model comparison depend on the choice of parameter
priors. 
}
{Our results confirm our earlier conclusion that the CMFs of the two Hi-GAL regions studied here 
have very similar shapes but different mass scales. Furthermore, the lognormal model appears 
to better describe the CMF measured in the two Hi-GAL regions studied here. However, this 
preliminary conclusion is dependent on the choice of parameters priors. 
}



\maketitle

\section{Introduction}
\label{sec:intro}

Stars form in dense, dusty cores, or clumps, of molecular clouds, but the
physical processes that regulate the transition from molecular clouds/clumps
to (proto)stars are still being debated.
In particular, the relationship between the mass distribution of molecular clumps
(also known as the ``core (or clump) mass function'', or CMF)
and the stellar initial mass function (IMF), is poorly understood \citep{mckee2007}.
In order to improve our understanding of this relationship, it is necessary
to undertake the study of statistically significant samples of pre- and proto-stellar
clumps.

The {\it Herschel} infrared GALactic Plane Survey (Hi-GAL),
a key program  of the {\it Herschel}
Space Observatory (HSO) to carry out a 5-band photometric imaging survey
at 70, 160, 250, 350, and $500\,\mu$m  of a $| b | \le 1^{\circ}$-wide strip of
the Milky Way Galactic plane \citep{molinari2010PASP}, is now providing us
with large samples of starless and proto-stellar cores, in a variety of
star-forming environments.
In the first paper (\citealp{olmi2013}, Paper I, herefater) we gave a
general description of the Hi-GAL data and described the source extraction and
photometry techniques. We also determined the spectral energy distributions 
and performed a statistical analysis of the CMF in the two regions mapped by HSO
during its science demonstration phase (SDP).
The two SDP fields were centered at $\ell=59^{\circ}$ and $\ell=30^{\circ}$  and the 
final maps spanned $\simeq 2^{\circ}$ in both Galactic longitude and latitude.

The goal of this second paper is twofold. On one side we build on the premises of Paper I, 
and apply a full Bayesian analysis to the CMF of the two SDP fields. First, we determine the
Bayesian posterior probability distribution of the parameters specific to each of the
CMF models analyzed in this work, powerlaw and lognormal. Next, we carry out a
quantitative comparison of these models, given data and an explicit set of assumptions.

The other major aim of our paper is to compare the results of 
several popular methods implementing Bayesian analysis, in order to highlight the effects 
that different algorithms may have on the results. In particular, we are interested in
analyzing the impact of the choice of {\it priors} and the influence of various constraints 
on the statistical conclusions for the preferred values of the parameters.

The outline of the paper is thus the following: in Section~\ref{sec:FittingModels}, 
we summarize the models used to describe the mass distribution.
In Section~\ref{sec:bayes}, we give a general description of Bayesian inference
and how it is applied to the analysis of the CMF.
We describe the algorithms used in Section~\ref{sec:globlike} and discuss our
results in Section~\ref{sec:discussion}. We finally draw
our conclusions in Section~\ref{sec:conclusions}.

\section{
Description of models used to fit the CMF }
\label{sec:FittingModels}

In the following sub-sections we give a short description of the mathematical functions
used in this analysis, but the reader should refer to Paper I for more details.


\subsection{Definitions}
\label{sec:defs}

We start by defining the CMF, $\xi(M)$, 
in the general case of a {\it continuous} distribution.
If ${\rm d}N$ represents the number of objects of mass $M$ lying between $M$ and $M + {\rm d}M$,
then we can define the number density distribution per mass interval, 
${\rm d}N/{\rm d}M$ with the relation \citep{chabrier2003}:
\begin{equation}
\xi(M) = \frac{{\rm d}N}{{\rm d}M} =  \frac{\xi(\log M)}{M \, \ln 10} =
\left( \frac{1}{M \, \ln 10} \right)  \frac{{\rm d}N}{{\rm d}\log M}
\label{eq:massfunct}
\end{equation}
thus, $\xi(M) {\rm d}M$ represents the number of objects with mass $M$ lying in the interval 
$\left[M, M+{\rm d}M \right]$.
The {\it probability} of a mass falling in the interval $\left[M, M+{\rm d}M \right]$ 
can be written for a continuous distribution as $p(M) {\rm d}M$, where $p(M)$ represents the
mass probability density function (or distribution, PDF). For the case of {\it discrete} data,
$p(M)$ can be written as:  
\begin{equation}
p(M) = \frac{\xi(M)}{N_{\rm tot}} 
\label{eq:PDF}
\end{equation}
where $N_{\rm tot}$ represents the total number of objects being considered in the sample. 
The PDF and CMF must obey the following normalization conditions (which we write here
for continuous data):
\begin{eqnarray}
& & \int_{M_{\rm inf}}^{M_{\rm sup}} p(M) {\rm d}M  =  1  \,\,\,  {\rm and} \nonumber \\
& & \int_{M_{\rm inf}}^{M_{\rm sup}} \xi(M) {\rm d}M  =  N_{\rm tot}
\label{eq:norm}
\end{eqnarray}
where $M_{\rm inf}$ and $M_{\rm sup}$ denote respectively the inferior and superior 
limits of the mass range for the objects in the sample, 
beyond which the distribution does not follow the specified behavior.

\subsection{Powerlaw form}
\label{sec:powerlaw}

The most widely used functional form for the CMF is the powerlaw: 
\begin{eqnarray}
\xi_{\rm pw}(\log\,M) & = & A_{\rm pw} \, M^{-\alpha},  \,\,\, {\rm or}  \\
\xi_{\rm pw}(M) & = &  \frac{A_{\rm pw}}{\ln 10}\,  M^{-\alpha-1}.
\label{eq:powerlaw}
\end{eqnarray}
where $A_{\rm pw}$ is the normalization constant.
The original Salpeter value for the IMF is $\alpha=1.35$ \citep{salpeter1955}.

The PDF of a powerlaw (continuous) distribution is given by \citep{clauset2009}:
\begin{equation}
p_{\rm pw}(M) = C_{\rm pw} \, M^{-\alpha-1}
\label{eq:PDFpw}
\end{equation}
where the normalization constant can be approximated as 
$C_{\rm pw}  \simeq \alpha \, M_{\rm inf}^{\alpha}$, if $\alpha > 0$ and 
$M_{\rm sup} \gg M_{\rm inf}$ (see Paper I and references therein).
As it will be described later in Section~\ref{sec:bayes}, Bayesian inference 
provides a technique to estimate the probability distribution of the model parameters
$\alpha$ and $M_{\rm inf}$.

\subsection{Lognormal form}
\label{sec:logn}

The continuous lognormal CMF can be written (e.g., \citealp{chabrier2003}):
\begin{eqnarray}
\xi_{\rm ln}(\ln\,M) = \frac{A_{\rm ln}}{\sqrt{2\pi} \, \sigma} \, 
\exp \left[ - \frac{(\ln M - \mu)^2}{2 \sigma^2} \right]
\label{eq:logn}
\end{eqnarray}
where $\mu$ and $\sigma^2 = \langle (\ln M - \langle \ln M \rangle )^2 \rangle$
denote respectively the mean mass and the variance in units of $\ln M$, and 
$A_{\rm ln}$ represents a normalization constant (see Paper I).

The PDF of a continuous lognormal distribution can be written as (e.g., \citealp{clauset2009}):
\begin{eqnarray}
p_{\rm ln}(M)  =  \frac{C_{\rm ln}}{M} \,
\exp \left[ - x^2 \right]
\label{eq:PDFln}
\end{eqnarray}
where we have defined the variable $x(M)=(\ln M - \mu)/(\sqrt{2} \sigma)$.
If the condition $M_{\rm sup} \gg M_{\rm inf}$
holds,  the normalization constant, $C_{\rm ln}$, can be approximated as (see Paper I):
\begin{equation}
C_{\rm ln} \simeq \sqrt{\frac{2}{\pi \sigma^2}  } \, \times
\left[ {\rm erfc}(x_{\rm inf}) \right]^{-1}
\label{eq:PDFCln2}
\end{equation}
where $x_{\rm inf} = x(M_{\rm inf})$.
As already mentioned for the powerlaw case, Bayesian inference will allow us to estimate 
the probability distribution of the three model parameters $\mu$, $\sigma$ and $M_{\rm inf}$.

\section{Bayesian Inference}
\label{sec:bayes}

\subsection{Overview of Bayesian methodology and prior information}
\label{sec:bayesintro}

Our main goal is to confront theories for the origin of the IMF with an analysis of the
CMF data that provide information on the processes responsible for cloud
fragmentation and clump formation. Bayesian inference allows the quantitative
comparison of alternative models, given the data and an explicit set of assumptions.
This last topic is known as {\it model selection}, i.e., the problem of distinguishing
competing models, generally featuring different numbers of free parameters.

The Bayesian statistics also provides a mathematically well-defined framework that allows
to determine the posterior probability distribution of the parameters of
a given model. As we have already seen in Section~\ref{sec:FittingModels}, the powerlaw model 
for the CMF depends on two parameters, $\alpha$ and $M_{\rm inf}$, while the lognormal model
contains three parameters, $\mu$, $\sigma$ and $M_{\rm inf}$ (unless $M_{\rm inf}$ is considered
a {\it fixed} parameter, see Section~\ref{sec:minfix}).


A very distinctive feature of 
Bayesian inference is that it deals quite naturally with {\it prior information} 
(for example, on the parameters of a given model), which in many cases is highly 
relevant, as for example when the parameters of interest have a physical meaning 
that restricts their possible values (e.g., masses, or positive quantities in general).
The prior choice in Bayesian statistics has been regarded both as a weakness and as a 
strength. In principle, prior assignment eventually becomes irrelevant as better and 
better data make the posterior distribution of the parameters dominated by the likelihood
of the data (see, e.g., \citealp{trotta2008a}). However, more often the data are not strong enough 
to override the prior, in which case the final inference may depend on the prior choice.
If different prior choices lead to different posteriors one should conclude that the data 
are not {\it informative} enough to completely override our prior state of knowledge.
An analysis of the role of priors in cosmological parameter extraction and Bayesian 
cosmological model building has already been presented by \citet{trotta2008b}.

The situation is even more critical in model selection. In this case, the impact of the 
prior choice is much stronger, and care should be exercised in assessing how much 
the outcome would change for physically reasonable changes in the prior (see, e.g., 
\citealp{berger2001} and \citealp{pericchi2005}). 
In addition to being nonrobust with respect to the choice of parameters priors,
Bayesian model selection also suffers from another deep difficulty, specifically 
the computation of a quantity, the {\it global likelihood}, which is difficult
to calculate to the required accuracy.

In this section we will first give a short introduction to Bayesian inference,
by reviewing the basic terminology and describing the most common prior types.
We also briefly discuss model comparison and define the global likelihood. 
The mathematical tools required to efficiently evaluate the global likelihood
and their limitations will then be discussed in the subsequent sections.

\subsection{Definitions}
\label{sec:bayesdef}

Here we give a short list of defintions that will be used later.

\begin{enumerate}

\item We denote a particular model by the letter ${\mathcal M}$. This particular
model is characterized by $Q$ parameters, which we denote by $\theta_{q}$, $q=1,...,Q$
(with size $Q$ dependent on the model).
The set of $\theta_{q}$ constitutes the parameter vector {\boldmath $\theta$}.
In this paper we consider two models for the CMF: the powerlaw ($Q=2$, 
$\mbox{{\boldmath $\theta$}} = [\alpha, M_{\rm inf}]$) and the lognormal
($Q=3$, $\mbox{{\boldmath $\theta$}} = [\mu, \sigma, M_{\rm inf}]$) models.

\item We denote the data by the letter $D$. In this work the data consist of
the observed CMF for the $\ell=59^{\circ}$ and $\ell=30^{\circ}$ Hi-GAL SDP fields (see Paper I).
More restrictive selection criteria have been applied to ensure that all methods described
in Section~\ref{sec:globlike} did converge.
As already done in the previous sections, individual clump masses will be denoted by
$M_i$ (not to be confused with the model, ${\mathcal M}$).

\item In the following we will use the {\it likelihood} of a given set of data $D$,
i.e. the combined probability that $D$ would be obtained from model ${\mathcal M}$
and its set of parameters $\theta_{q}$, which we denote by
$P(D | \mbox{{\boldmath $\theta$}}, {\mathcal M})$ or ${\mathcal L}$.  For the
present case and for a set of data $D=\{M_i\}$, the likelihood can be written as:
\begin{equation}
P(D | \mbox{{\boldmath $\theta$}}, {\mathcal M}) = {\mathcal L}(\mbox{{\boldmath $\theta$})} =
\prod_{i=1}^{N_{\rm tot}} \, p(M_i; \mbox{{\boldmath $\theta$})}
\label{eq:likelihooddef}
\end{equation}
where it is assumed that the data are drawn from the PDF associated with
model ${\mathcal M}$, denoted by $p(M_i; \mbox{\boldmath $\theta$})$
(see Section~\ref{sec:FittingModels}).

\item In principle, the model parameters $\theta_{q}$ can take any value, unless
we have some information limiting their range.
We can constrain the expected ranges of parameter values by assigning
the probability distributions of the unknown parameters $\theta_{q}$.
These are called the {\it parameters prior probability distribution}
(often called simply the {\it parameters prior} 
and are denoted by $P(\theta_q | {\mathcal M})$. Hence:
\begin{equation}
P( \mbox{{\boldmath $\theta$}} | {\mathcal M}) =
\prod_{q=1}^Q \, P(\theta_q | {\mathcal M})
\end{equation}

\item
In contrast to traditional point estimation methods (e.g., 
maximum-likelihood estimation, or MLE)
Bayesian inference does not provide specific estimates for
the parameters. Rather, it provides a technique to estimate the {\it probability
distribution} (assumed to be continuous) of each model parameter $\theta_{q}$,
also known as the {\it posterior} PDF, or simply the posterior distribution, 
$P(\mbox{{\boldmath $\theta$}} | D, {\mathcal M})$. In Bayesian statistics, 
the posterior distribution encodes the full information coming from
the data and the prior, and it is given by the {\it Bayes theorem}:
\begin{equation}
P(\mbox{{\boldmath $\theta$}} | D, {\mathcal M}) =
\frac{ P(\mbox{{\boldmath $\theta$}} | {\mathcal M}) \,
P( D | \mbox{{\boldmath $\theta$}}, {\mathcal M}) }
{ {\mathcal P}( D | {\mathcal M})    } =
\frac{ P(\mbox{{\boldmath $\theta$}} | {\mathcal M}) \,
{\mathcal L}(\mbox{{\boldmath $\theta$})}   }
{ {\mathcal P}( D | {\mathcal M})    }
\label{eq:bayesth}
\end{equation}
where ${\mathcal P}( D | {\mathcal M}) $ is a
normalization factor and is often called the {\it global likelihood} or the {\it evidence}
for the model: 
\begin{equation}
{\mathcal P}( D | {\mathcal M}) = \int  
P( \mbox{{\boldmath $\theta$}} | {\mathcal M}) \, 
{\mathcal L}(\mbox{{\boldmath $\theta$})} \, 
{\rm d}\mbox{{\boldmath $\theta$}} 
\label{eq:globlik}
\end{equation}
thus the global likelihood of a model is equal to the weighted
(by the parameters prior, $P( \mbox{{\boldmath $\theta$}} | {\mathcal M})$)
average likelihood for its parameters.
We will be working mostly with logarithmic probabilities, thus Eq.~(\ref{eq:bayesth}) 
becomes:
\begin{equation}
\ln P(\mbox{{\boldmath $\theta$}} | D, {\mathcal M}) = const. + \ln P( \mbox{{\boldmath $\theta$}} | {\mathcal M}) +
\ln {\mathcal L}(\mbox{{\boldmath $\theta$})}
\label{eq:bayesthlog}
\end{equation}
where
\begin{equation}
\ln {\mathcal L}(\mbox{{\boldmath $\theta$})} = \ln \prod_{i=1}^{N_{\rm tot}} \, p(M_i; \mbox{{\boldmath $\theta$})} =
\sum_{i=1}^{N_{\rm tot}} \, \ln p(M_i; \mbox{{\boldmath $\theta$})} \, .
\label{eq:greg3}
\end{equation}

\end{enumerate}

\subsection{Specifying the parameter priors}
\label{sec:priors}

If we have some expectation of the ranges in which the parameter values lie,
then we can incorporate this information in the parameters priors
$P( \mbox{{\boldmath $\theta$}} | {\mathcal M})$. Even when parameters
in a given range of values are equally probable, we can specify 
plausible bounds on parameters.


Here we briefly introduce the two forms of priors most commonly used, i.e.,  
the {\it uniform} and {\it Jeffreys'} priors, then in Section~\ref{sec:globlike}
we will discuss the priors actually used in our computations.

\begin{enumerate}

\item
When dealing with {\it scale parameters} the preferred form is the Jeffreys' priors,
which assign equal probability per decade interval (appropriate for quantities
that are scale invariant), and are given by (see, e.g., \citealp{gregory2005}):
\begin{equation}
P( \theta_{\rm q} | {\mathcal M} ) =
\left \{ \begin{array}{l}
\frac{1}{\theta_{\rm q} \,\ln(\theta_{\rm q}^{\rm max} / \theta_{\rm q}^{\rm min} )}
, \hspace{0.2cm} {\rm for}   \hspace{0.2cm}
\theta_{\rm q}^{\rm min} \leq \theta_{\rm q} \leq \theta_{\rm q}^{\rm max} \\
0, \hspace{2.4cm} {\rm otherwise} \\
\end{array}
\right.
\label{eq:jeffreys}
\end{equation}
where $ [\theta_{\rm q}^{\rm min}, \theta_{\rm q}^{\rm max}]$
represents the range allowed for parameter $\theta_{\rm q}$ to vary.

\item
On the other hand, when dealing with {\it location parameters} the preferred
form is the uniform priors, which give uniform probability per arithmetic interval:
\begin{equation}
P( \theta_{\rm q} | {\mathcal M} ) =
\left \{ \begin{array}{l}
\frac{1}{\theta_{\rm q}^{\rm max} - \theta_{\rm q}^{\rm min} }, \hspace{0.2cm} {\rm for}   \hspace{0.2cm}\theta_{\rm q}^{\rm min} \leq \theta_{\rm q} \leq \theta_{\rm q}^{\rm max} \\
0, \hspace{1.5cm} {\rm otherwise.} \\
\end{array}
\right.
\label{eq:uniform}
\end{equation}

\end{enumerate}

\subsection{Model comparison}
\label{sec:ModComp}

In many cases, such as the present one, more than one parameterized model is available
to explain a given set of data, and it is thus of interest to compare them. The models
may differ in form and/or in number of parameters. Use of Bayes' theorem allows to
compare competing models by calculating the probability of each model as a whole.
The equivalent form of Eq.~(\ref{eq:bayesth}) to calculate the posterior probability of
a model ${\mathcal M}$, $P({\mathcal M} | D, I)$, which represents the
probability that model ${\mathcal M}$ has actually generated the data $D$,
is the following \citep{gregory2005}:
\begin{equation}
P({\mathcal M} | D, I) = \frac{
P( {\mathcal M} | I) \, {\mathcal P}( D | {\mathcal M}, I) }
{ P( D | I) } 
\label{eq:bayesthmod}
\end{equation}
where $I$ represents our prior information that one of the models under consideration
is true. One can recognize ${\mathcal P}( D | {\mathcal M}, I)$ as the 
global likelihood for model ${\mathcal M}$, which can be calculated according to 
Eq.~(\ref{eq:globlik}).  $P( {\mathcal M} | I)$ represents the 
{\it prior probability} for model ${\mathcal M}$, while the term at the
denominator $P( D | I)$ is again a normalization constant, obtained by summing the products
of the priors and the global likelihoods of all models being considered.


Then, the plausibility of two different models ${\mathcal M_1}$ and ${\mathcal M_2}$,
parameterized by the model parameters vectors {\boldmath $\theta$}$_1$ and
{\boldmath $\theta$}$_2$, can be assessed by the {\it odds ratio}:
\begin{equation}
\frac{P({\mathcal M_2} | D, I) }{P({\mathcal M_1} | D, I)} = BF_{21} \,
\frac{P({\mathcal M_2}, I) }{P({\mathcal M_1}, I) }
\label{eq:oddsratio}
\end{equation}
which can be interpreted as the ``odds provided by the data for model ${\mathcal M_2}$
versus ${\mathcal M_1}$ ''.  In Eq.~(\ref{eq:oddsratio}) we have also introduced
the {\it Bayes factor} (e.g., \citealp{gelfand1994}, \citealp{gregory2005}):
\begin{equation}
BF_{21} = \frac{ {\mathcal P}( D | {\mathcal M_2}, I) }{ {\mathcal P}( D | {\mathcal M_1}, I) } =
\frac{ \int 
P( \mbox{{\boldmath $\theta$}$_2$} | {\mathcal M_2}) \, {\mathcal L}(\mbox{{\boldmath $\theta$}$_2$)} \, 
{\rm d}\mbox{{\boldmath $\theta$}$_2$} }
{ \int 
P(\mbox{{\boldmath $\theta$}$_1$} | {\mathcal M_1}) \, {\mathcal L}(\mbox{{\boldmath $\theta$}$_1$)} \, 
{\rm d}\mbox{{\boldmath $\theta$}$_1$}  }
\label{eq:bayesgen}
\end{equation}
In general it is also assumed that $P({\mathcal M_2}, I) = P({\mathcal M_1}, I)$
(i.e., no model is favoured over the other),
thus the odds ratio becomes equal to the Bayes factor.
A value of $BF_{21} > 1$ would indicate that the data provide evidence in favor of 
model ${\mathcal M_2}$ vs. the alternative model ${\mathcal M_1}$.
Usually, the Bayes factor is quoted in $\ln$ units,
i.e. we change to $ \ln (BF)$:
\begin{eqnarray}
{\mathcal B}{\mathcal F}_{21} = \ln (BF)_{21} & = & \ln {\mathcal P}( D | {\mathcal M_2}, I) -
\ln {\mathcal P}( D | {\mathcal M_1}, I) 
\label{eq:bayespwln}
\end{eqnarray}
Then, $sgn({\mathcal B}{\mathcal F})$ will indicate the most probable model, with
positive values of ${\mathcal B}{\mathcal F}$ favoring model ${\mathcal M_2}$
and negative values favoring model ${\mathcal M_1}$.

Thus, for each model we must compute the global likelihoods, which means evaluating
the integrals in Eq.~(\ref{eq:bayesgen}),  which can be written  in general
as\footnote{Here and in the following, we will keep using
the symbol of simple integration, $\int$, instead of that for multiple integration over
the parameters space, $\int ... \int$.  }:
\begin{eqnarray}
\label{eq:MargLike}
& &  {\mathcal P}( D | {\mathcal M_j}, I)  =
\int {\mathcal L}(\mbox{{\boldmath $\theta$}$_j$)} \,
\, P( \mbox{{\boldmath $\theta$}$_j$} | {\mathcal M_j}) \,
 {\rm d}\mbox{{\boldmath $\theta$}$_j$}  =     \nonumber \\
& & = \int \left [ \prod_{i=1}^{N_{\rm tot}} \, p_{\rm j}(M_i; \mbox{{\boldmath $\theta$}$_{\rm j}$)} \right ]
\, \left [ \prod_{q=1}^{Q} \, P( \theta_{\rm jq} | {\mathcal M_j} )     \right ] \,
{\rm d}\mbox{{\boldmath $\theta$}$_j$} \\
& &  j=1,2 \hspace{5mm} q=1,...,Q_j \nonumber
\end{eqnarray}
which can be written more explicitly, for example in the powerlaw case and using 
Jeffreys priors for both parameters, as:
\begin{eqnarray}
\label{eq:MargLikePW1}
& &  {\mathcal P}( D | {\mathcal M_{\rm pw}}, I)  =  \nonumber \\
& & = \int  \left [ \prod_{i=1}^{N_{\rm tot}} \, C_{\rm pw} \, M_i^{-\alpha} \right ] \,
P( \mbox{{\boldmath $\theta$}$_{\rm pw}$} | {\mathcal M_{\rm pw}}) \,
{\rm d}\alpha \,\, {\rm d}M_{\rm inf},  \hspace{3mm} {\rm with}  \\
& & P( \mbox{{\boldmath $\theta$}$_{\rm pw}$} | {\mathcal M_{\rm pw}}) =
\frac{1}{M_{\rm inf} \, \ln(M_{\rm inf}^{\rm max}/M_{\rm inf}^{\rm min} )} \times \nonumber \\
& & \frac{1}{\alpha \, \ln(\alpha_{\rm max}/ \alpha_{\rm min} ) } \nonumber
\end{eqnarray}
where we have used Eq.~(\ref{eq:PDFpw}) and $C_{\rm pw}$ has been given in 
Section~\ref{sec:powerlaw}.

\section{Computation of model parameters and global likelihood}
\label{sec:globlike}

We now turn to the description of various methods that allow the computation
of the posterior distributions of the model parameters which, in some cases,
also allow to estimate the global likelihood as a by-product.
Severeal open software resources exist that can perform the computation
of model parameters. However, estimating the multi-dimensional integrals in 
equations of the type (\ref{eq:MargLike}) and (\ref{eq:MargLikePW1})
is impossible to be done analytically in most cases, and is
otherwise computationally very intensive when done numerically. 
Therefore, popular statistical packages (e.g., 
WinBUGS\footnote{http://www.mrc-bsu.cam.ac.uk/bugs/winbugs/contents.shtml}, 
${\mathcal R}$\footnote{http://www.r-project.org/}) are usually not able to
compute the global likelihood directly, and more specialized programs
or {\it ad-hoc} procedures must be used.

As it will be discussed in greater detail in Section~\ref{sec:discussion}, we note 
that all methods described below have proved to be sensitive, to various degrees,
to the priors type and range as well as to other parameters specific to each algorithm.
In addition, some of these algorithms would either crash, if for example
the prior range was too wide, or some of the model parameters (e.g., $\alpha$, $\mu$)
would converge toward one end of the prior range.

Therefore, in order for the comparison of the various methods to be meaningful, 
and in order to avoid software problems, we selected the same type of priors and range,
and also an even more restrictive sub-set of our data, 
compared to those used in Paper I.  In particular, we selected for our comparison
{\it uniform} priors (see Table~\ref{tab:priors}), since only {\it proper} priors 
(i.e., uniform and Gaussian) were immediately implementable in WinBUGS. In addition, uniform
priors are {\it non-informative} (i.e., supposedly provides ``minimal'' influence
on the inference) compared to Gaussian priors.  

%
\begin{table}
\caption{Ranges for uniform priors used
toward the  $\ell=30^{\circ}$ and $\ell=59^{\circ}$ fields.
}
\label{tab:priors}
\centering
\begin{tabular}{lcccccc}
\hline\hline
Region   & \multicolumn{2}{c}{Powerlaw} & \multicolumn{4}{c}{Lognormal}  \\
\cline{2-3}
\cline{5-7}
                     & $\alpha$  & $M_{\rm inf}$  &  & $\mu$             & $\sigma$          & $M_{\rm inf}$  \\
                     &           & [$M_\odot$]    &  & [$\ln(M_\odot)$]  & [$\ln(M_\odot)$]  & [$M_\odot$]    \\
\hline
$\ell=30^{\circ}$    & [0,3]     & [10,30]        &  & [0.8,20]          & [0.4,10]          & [10,30]        \\
$\ell=59^{\circ}$    & [0,3]     & [0.3,1.5]      &  & [0.2,5]           & [0.2,4]           & [0.3,0.7]      \\
\hline
\end{tabular}
\end{table}


\subsection{Laplace approximation and harmonic mean estimator }
\label{sec:WBUGS}

In this section we describe two methods to implement the
computation of the global likelihood that use open software resources,
and thus do not require the development by the user of novel 
specific software.

%
%
%
%
%
%
%
%
%
%
%
%
%
%
%
\begin{table*}
\caption{
Mean values and standard deviations estimated from the posterior
distributions of the parameters,
obtained using the methods described in the text (WinBUGS, MHPT and MultiNest), for the  powerlaw distribution.
Parameter $M_{\rm inf}$ is undetermined by all methods considered. Therefore in the case of the Laplace method
(requiring the posterior means) the mid value of the priors range was used.
Only results obtained using {\it uniform} priors are shown (see Table~\ref{tab:priors}).
}
\label{tab:power}
\centering
\begin{tabular}{lcccccr}
\hline\hline
%
Method                   &  & \multicolumn{2}{c}{$\ell=30^{\circ}$}           &    & \multicolumn{2}{c}{$\ell=59^{\circ}$}    \\
\cline{3-4}
\cline{6-7}
                         &  & $\alpha$  & $\ln{\mathcal P}(D|{\mathcal M})$   &    & $\alpha$  & $\ln{\mathcal P}(D|{\mathcal M})$  \\
\hline
HME                      &  & $0.73\pm0.03$    & $-9008$                      &    & $0.56\pm0.02$    & $-1924$                     \\
Laplace                  &  & $0.73\pm0.03$    & $-9454$                      &    & $0.56\pm0.02$    & $-2051$                     \\
MHPT                     &  & $0.7\pm0.1$      & $-4753$                      &    & $0.55\pm0.01$    & $-4985$                     \\
MultiNest                &  & $0.73\pm0.02$    & $-9018$                      &    & $0.56\pm0.02$    & $-1932$                     \\
\hline
\end{tabular}
%
%
\end{table*}

\subsubsection{Laplacian approximation}
\label{sec:laplace}

One of the most popular approximation of the global likelihood is the
so called Laplace approximation, which results in (e.g., \citealp{gregory2005}, \citealp{ntzoufras2009}):
\begin{equation}
{\mathcal P}( D | {\mathcal M}) \approx (2\pi)^{Q/2} \,
|\mbox{\boldmath $H$}(\hat{\mbox{\boldmath $\theta$}}) |^{-1/2} \,
P(D | \hat{\mbox{\boldmath $\theta$}}, {\mathcal M}) \,
P( \hat{\mbox{\boldmath $\theta$}} | {\mathcal M})
\label{eq:laplace_one}
\end{equation}
where $Q$ represents the number of parameters (see Section~\ref{sec:bayesdef}, item 1),
$\hat{\mbox{\boldmath $\theta$}}$ is the posterior mode of the parameters
of model $ {\mathcal M}$, and $\mbox{\boldmath $H$}$ is equal to the minus of the
second derivative matrix (with respect to the parameters) of the posterior PDF, i.e.,
$ P(\mbox{{\boldmath $\theta$}} | D, {\mathcal M})$ (see Section~\ref{sec:bayesdef}, item 5),
evaluated at the posterior mode $\hat{\mbox{\boldmath $\theta$}}$.

As described by \cite{ntzoufras2009} the Laplace-Metropolis estimator can be used to
evaluate Eq.~(\ref{eq:laplace_one}), where $\hat{\mbox{\boldmath $\theta$}}$ and
$\mbox{\boldmath $H$}$ can be estimated from the output of a Markov Chain Monte Carlo
(MCMC) algorithm (see Section~\ref{sec:MCMC}). Thus, Eq.~(\ref{eq:laplace_one}) becomes:
\begin{eqnarray}
 \ln {\mathcal P}( D | {\mathcal M}) & \approx &  \frac{1}{2} Q \ln(2\pi) +
 \frac{1}{2} \ln |\mbox{\boldmath $R$}_\theta| +
\sum_{q=1}^Q \ln s_q + \nonumber \\
& & + \sum_{i=1}^{N_{\rm tot}} \ln p(M_i; \bar{\mbox{\boldmath $\theta$}}) +
\ln P( \bar{\mbox{\boldmath $\theta$}} | {\mathcal M})
\label{eq:laplace_two}
\end{eqnarray}
where the posterior means (replacing the posterior modes) of the parameters
of interest are denoted by $\bar{\mbox{\boldmath $\theta$}}$,
$\mbox{\boldmath $R$}_\theta$ represents the posterior correlation between
the parameters of interest, $s_q$ are the posterior standard deviations
of the $\theta_q$ parameters of model $ {\mathcal M}$, and
$p(M_i; \bar{\mbox{\boldmath $\theta$}})$ is the PDF associated to model $ {\mathcal M}$
and evaluated at data point $i$ (see Section~\ref{sec:bayesdef}, item 3).

Following \cite{ntzoufras2009} we estimate the posterior means, standard deviations and
correlation matrix from an MCMC run in WinBUGS, and then
we calculate the global likelihood from Eq.~(\ref{eq:laplace_two})  
in an external software such as ${\mathcal R}$, after importing the posterior summaries and
the data. The results are listed in tables~\ref{tab:power} and
\ref{tab:logn}, while the Bayes factors are listed in Table~\ref{tab:globlike} and
will be discussed later in Section~\ref{sec:discussion}.

\begin{table*}
\caption{
Same as Table~\ref{tab:power} for the lognormal distribution.
}
\label{tab:logn}
\centering
\begin{tabular}{lcccccccr}
\hline\hline
Method                   &  & \multicolumn{3}{c}{$\ell=30^{\circ}$}         &    & \multicolumn{3}{c}{$\ell=59^{\circ}$}    \\
\cline{3-5}
\cline{7-9}
             &  & $\mu \,$[$\ln M_\odot$]  & $\sigma \,$[$\ln M_\odot$]  & $\ln{\mathcal P}(D|{\mathcal M})$    &   
                & $\mu \,$[$\ln M_\odot$]  & $\sigma \,$[$\ln M_\odot$]  & $\ln{\mathcal P}(D|{\mathcal M})$        \\
\hline
HME          &  & $4.76\pm0.02$            & $0.83\pm0.06$             & $-6424$                              &   
                & $1.43\pm0.04$            & $1.29\pm0.15$             & $-543$                               \\  
Laplace      &  & $4.76\pm0.02$            & $0.83\pm0.06$             & $-9130$                              &   
                & $1.43\pm0.04$            & $1.29\pm0.15$             & $-1892$                              \\  
MHPT         &  & $4.6\pm0.1$              & $1.1\pm0.1$               & $-4610$                              &   
                & $1.28\pm0.01$            & $1.17\pm0.01$             & $-4622$                              \\  
MultiNest    &  & $4.0\pm0.1$              & $1.4\pm0.1$               & $-8904$                              &   
                & $1.1\pm0.1$              & $1.3\pm0.1$               & $-1823$                              \\  
\hline
\end{tabular}
%
%
\end{table*}

\subsubsection{Harmonic mean estimator}
\label{sec:HME}

The harmonic mean estimator (HME) is also based on an MCMC run and 
provides the following estimate for the global likelihood \citep{ntzoufras2009}:
\begin{equation}
{\mathcal P}( D | {\mathcal M}) \approx \left \{   
\frac{1}{T} \sum_{\rm t=1}^T {\mathcal L}(\mbox{{\boldmath $\theta$}$_t$)}^{-1}
\right \}^{-1}   
\label{eq:HME}
\end{equation}
%
where ${\mathcal L}(\mbox{{\boldmath $\theta$}$_t$)}$ represents the likelihood
of the data corresponding to the $t-$th run of the MCMC simulation, having
a total of $T$ samples. Although very simple, this estimator is quite unstable
and sensitive to small likelihood values and hence it is not recommended. However,
in Table~\ref{tab:globlike} we present the HME values obtained by us with WinBUGS 
as a comparison for the Laplace-Metropolis method.

\subsection{Computation of the global likelihood with the Metropolis-Hastings algorithm}
\label{sec:MH}

In this section we describe how the global likelihood and the Bayes factor can
also be estimated by implementing our own MCMC procedure.

\subsubsection{Markov Chain Monte Carlo}
\label{sec:MCMC}


As we previously mentioned in Section~\ref{sec:bayesdef}, in Bayesian inference
parameter estimation consists of calculating the posterior PDF, or density,
$P(\mbox{{\boldmath $\theta$}} | D, {\mathcal M})$, given by Eq.~(\ref{eq:bayesth}).
However, since we must vary all $\theta_q$ parameters, we need a method for
exploring the parameter space, because gridding in each
parameter direction would lead to an unmanageably large number of sampling
points.
The $Q-$dimensional parameter space can be explored with the aid of MCMC techniques,
which are able to draw samples from the unknown posterior density
(also called the {\it target distribution}) by constructing a pseudo-random walk
in model parameter space, such that the number of samples drawn from a particular
region is proportional to its posterior density.
Such a pseudo-random walk is achieved by generating a {\it Markov chain}, which
we create using the {\it Metropolis-Hastings} (MH) algorithm
(Metropolis et al. 1953, Hastings 1970).

Briefly, the MH algorithm proceeds as follows. Given the posterior density
$P(\mbox{{\boldmath $\theta$}} | D, {\mathcal M})$ and any starting position
$\mbox{{\boldmath $\theta$}$_{\rm t}$}$ in the parameter space, the step to the next position
$\mbox{{\boldmath $\theta$}$_{\rm t+1}$}$ in the random walk is obtained
from a {\it proposal distribution} (for example, a normal distribution)
$g( \mbox{{\boldmath $\theta$}$_{\rm t+1}$} | \mbox{{\boldmath $\theta$}$_{\rm t}$})$.
Assuming $g$ to be symmetric in $\mbox{{\boldmath $\theta$}$_{\rm t}$}$ and
$\mbox{{\boldmath $\theta$}$_{\rm t+1}$}$, the requirement of detailed balance leads
to the following rule: accept the proposed move to $\mbox{{\boldmath $\theta$}$_{\rm t+1}$}$
if the {\it Metropolis ratio} $r\ge1$, where
$r = P(\mbox{{\boldmath $\theta$}$_{\rm t+1}$} | D, {\mathcal M}) /
P(\mbox{{\boldmath $\theta$}$_{\rm t}$} | D, {\mathcal M})   $.
If $r < 1$, remain at $\mbox{{\boldmath $\theta$}$_{\rm t}$}$.
This sequence of
proposing new steps and accepting or rejecting these steps is then iterated
until the samples (after a burn-in phase) have converged to the target distribution.
Since in the Metropolis ratio the factor at the denominator of Eq.~(\ref{eq:bayesth})
cancels out, then the evaluation of $r$ requires only the calculation of the
parameters priors and of the likelihoods, but {\it not} of the global likelihood,
${\mathcal P}( D | {\mathcal M}) $.

A modified version of the MH algorithm to fully explore all regions in parameter
space containing significant probability employs the so-called {\it parallel tempering}
(MHPT; \citealp{gregory2005}, see also \citealp{handberg2011}). 
In the MHPT method several versions, or {\it chains} ($n_\beta$ in total),
of the MH algorithm are launched in parallel, thus generating  a discrete set
of progressively flatter versions of the target distribution, also known as the
{\it tempered distributions}. Each of these $n_\beta$ chains
is characterized by a different tempering parameter, $\beta$, and the new
target distributions can be written by modifying Eq.~(\ref{eq:bayesth}):
\begin{equation}
P(\mbox{{\boldmath $\theta$}} | D, {\mathcal M}, \beta) \propto
 P(\mbox{{\boldmath $\theta$}} | {\mathcal M}) \,
{\mathcal L}(\mbox{{\boldmath $\theta$})}^\beta, \hspace{0.5cm} 0 < \beta \leq 1 
\label{eq:posteriorbeta}
\end{equation}
For $\beta = 1$, we recover the target distribution.
The MHPT method allows to visit regions of parameter space containing significant
probability, not accessible to the basic algorithm. 
The  main steps of the Metropolis-Hastings algorithm, with the inclusion of parallel 
tempering, are described in Appendix~\ref{sec:procMHPT}.

\subsubsection{Application of MHPT method to model comparison}
\label{sec:modcomp}

Going back, now, to the issue of model comparison, an important property
of the MHPT method is that samples drawn from the tempered distributions can
be used to compute the global likelihood, ${\mathcal P}( D | {\mathcal M}) $, of a given
model ${\mathcal M}$. In fact, it can be shown that the global log-likelihood of a
model is given by (for a derivation see \citealp{gregory2005}):
\begin{equation}
\ln {\mathcal P}( D | {\mathcal M}) = \int_0^1 \langle \ln 
{\mathcal L}(\mbox{{\boldmath $\theta$})} \rangle_\beta \, {\rm d}\beta
\label{eq:greg1}
\end{equation}
where
\begin{equation}
\langle \ln {\mathcal L}(\mbox{{\boldmath $\theta$})} \rangle_\beta =
\frac{1}{T} \sum_{t=1}^T \ln {\mathcal L}(\mbox{{\boldmath $\theta$}$_{t\beta}$)}
\label{eq:greg2}
\end{equation}
where $T$ is the number of samples in each set after the burn-in period.
The log-likelihoods in Eq.~(\ref{eq:greg2}),
$\ln {\mathcal L}(\mbox{{\boldmath $\theta$}$_{t\beta}$)}$, can be evaluated
using Eq.~(\ref{eq:likelihooddef}) and
from the MHPT results, which consist of sets of \{$ \mbox{\boldmath $\theta$}_t$\}
samples, one set (i.e., Markov chain, $\mbox{{\boldmath $\theta$}}_1 \rightarrow
\mbox{{\boldmath $\theta$}}_2 \rightarrow ... \rightarrow \mbox{{\boldmath $\theta$}}_t
\rightarrow ...$) for each value of the tempering parameter $\beta$.

As a by-product of the computation of the global likelihood,
the MHPT method can also be used to determine the posteriors of the 
parameters.  The results, with both the posterior summaries and the global likelihoods,
are also listed in tables~\ref{tab:power} and \ref{tab:logn}.

\subsection{Computation of the global likelihood using the nested sampling method} 
\label{sec:MT}

The main problem of the methods outlined in Section~\ref{sec:WBUGS} 
is the approximations involved, whereas the MCMC sampling methods, such as the MHPT 
technique described in Section~\ref{sec:MH}, may have problems in estimating the 
parameters of some model, if the resulting posterior distribution is for example 
multimodal.  In addition, calculation of the Bayesian evidence for each model is still 
computationally expensive using MCMC sampling methods.

The nested sampling method introduced by \citet{skilling2004}, is supposed to greatly 
reduce the computational expense of calculating evidence and also produces posterior 
inferences as a by-product. This method replaces the multi-dimensional integral in 
Eq.~(\ref{eq:globlik}) with a one-dimensional integral over unit range:
\begin{equation}
{\mathcal P}( D | {\mathcal M}) = \int_0^1
{\mathcal L}(\mbox{{\boldmath X})} \,
{\rm d}\mbox{{\boldmath X}}
\end{equation}
where ${\rm d}\mbox{{\boldmath $X$}} = P( \mbox{{\boldmath $\theta$}} | {\mathcal M}) \,
{\rm d}\mbox{{\boldmath $\theta$}}$ is the element of ``prior volume''. A sequence 
of values $\mbox{{\boldmath $X_j$}}$ can then be generated and the evidence is
approximated numerically using standard quadrature methods (see \citealp{skilling2004}
and \citealp{feroz2008}). Here, we use the ``multimodal nested sampling algorithm'' 
(MultiNest\footnote{http://ccpforge.cse.rl.ac.uk/gf/project/multinest/}, 
\citealp{feroz2008}, \citealp{feroz2009}) to calculate both global likelihood and posterior
distributions.  The results are summarized in tables~\ref{tab:power} and \ref{tab:logn}.

\section{Discussion}
\label{sec:discussion}

We now turn to the discussion of the effects of priors and different algorithms   
on the CMF parameter inference and model comparison using Bayesian statistics.
As it was mentioned in Section~\ref{sec:globlike}, our comparison is 
limited to uniform priors only, because they are non-informative and also
because of other software constraints. 

Our purpose is not to perform a general analysis of the impact of priors type and range
on Bayesian inference since, as previously discussed in Section~\ref{sec:bayesintro},
this is a very complex topic that goes well beyond
the scopes of the present work. Instead, we were interested for the two
distributions considered here in analyzing the sensitivity of our results to
some user-specified constraints. In particular, we were interested in the role
of the parameter $M_{\rm inf}$ which is clearly critical for both powerlaw and lognormal
distributions, as discussed below.

\subsection{Powerlaw results}
\label{sec:compPW}

\subsubsection{Non-regular likelihood: considering $M_{\rm inf}$ a free parameter}
\label{sec:minfree}

We start our discussion by analyzing the results of the posterior distributions
for the powerlaw model, when the parameter $M_{\rm inf}$ is free to vary. Then,
Table~\ref{tab:power} shows 
that the three methods described in Section~\ref{sec:globlike} deliver 
remarkably similar values of the $\alpha$ parameter, separately for the two SDP fields, 
and with the prior ranges shown in Table~\ref{tab:priors}.
However, the powerlaw slope estimated for the two fields is different, and is also 
somewhat different from the values quoted in Paper I. 

%
%
\begin{figure}
%
%
\centering
\includegraphics[width=9.2cm,angle=0]{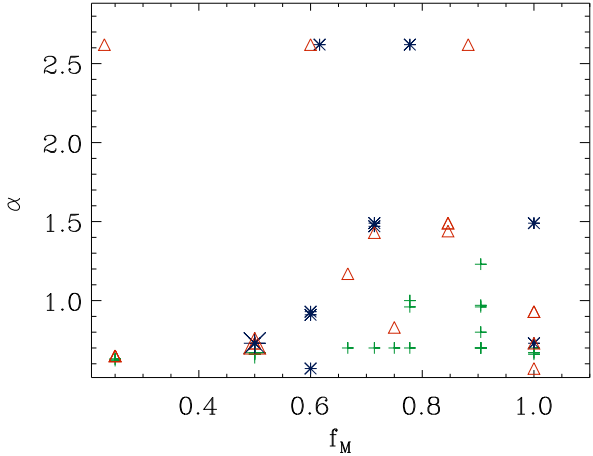}
\caption[ ]{
Comparison of results obtained for the parameter $\alpha$ (powerlaw case)
in the $\ell=30^{\circ}$ field using the MHPT (green plus signs),
MultiNest (red triangles) and WinBUGS (blue asterisks)
methods, as a function of $f_M = (M_{\rm inf2}-M_{\rm inf1})/(M_{\rm inf1}+M_{\rm inf2})$,  where $M_{\rm inf1}$
and $M_{\rm inf2}$ are the extremes of the prior range for $M_{\rm inf}$ (see text).
The larger symbols correspond to the values listed in Table~\ref{tab:power}.
}
\label{fig:comparisonl30}
\end{figure}

%
%
\begin{figure}
%
%
%
\centering
\includegraphics[width=9.2cm,angle=0]{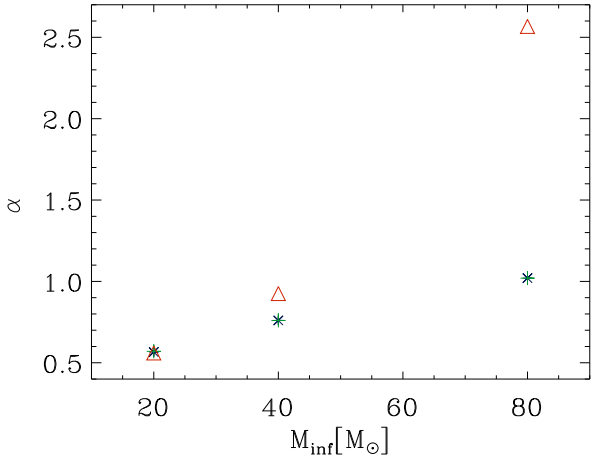}
\caption[ ]{
Comparison of results obtained for the parameter $\alpha$ (powerlaw case)
in the $\ell=30^{\circ}$ field keeping $M_{\rm inf}$ fixed. Symbols and colors are as
in Fig.~\ref{fig:comparisonl30}. The points representing the MHPT and WinBUGS
results overlap almost exactly, and the error bars on $\alpha$  are not shown because
they are tipically contained within the symbol size (see text).
}
\label{fig:comparisonl30Minfix}
\end{figure}

This discrepancy, however, is less significant compared to the sensitivity
of the posteriors on the priors range, and in particular the range 
$[M_{\rm inf1},M_{\rm inf2}]$ for the uniform prior on the parameter $M_{\rm inf}$. 
We checked this sensitivity toward one of the two SDP fields, the 
$\ell=30^{\circ}$ region. Thus, in Fig.~\ref{fig:comparisonl30} we plot the 
values of $\alpha$ obtained with the three methods discussed above, as a function of the 
parameter $f_M = (M_{\rm inf2}-M_{\rm inf1})/(M_{\rm inf1}+M_{\rm inf2})$.  
The parameter $f_M $ thus represents a measure of the amplitude of the prior range, and
the scatter in Fig.~\ref{fig:comparisonl30} is due either to the fact that different
$M_{\rm inf}$ ranges may have the same value of $f_M $ or also due to the variation of other
parameters specific to the method used. 
Despite their sensitivity to the parameter $f_M $, the values of $\alpha$
are much less sensitive to the range of its own uniform prior (see also Section~\ref{sec:minfix}). 

Looking at Fig.~\ref{fig:comparisonl30} it is not surprising that the values listed in
Table~\ref{tab:power} are somewhat different from those quoted in Paper I (Table 4). 
In fact, the values listed in Paper I could be easily reproduced with the proper choice 
of $f_M $. In addition, it should be noted that the values of $\alpha$ and $M_{\rm inf}$ 
listed in Paper I were determined using the PLFIT method \citep{clauset2009}, but even 
with this method the result for these parameters depends on whether an upper limit 
for $M_{\rm inf}$ is selected or not.

We also note that all of the methods used were unable to deliver a well defined 
value for the $M_{\rm inf}$ parameter, unless {\it Gaussian} (i.e., {\it informative}) 
priors were used. In fact, in all cases considered this parameter tends to converge 
toward the higher end of the prior range. However, this is not an effect caused by 
the specific data samples used. In fact, in order to test this issue
we generated a set of power-law distributed data, using the method described
in \citet{clauset2009}, and applied to it the MHPT and MultiNest methods. In both
cases $M_{\rm inf}$ tended to converge toward the higher end of the prior range.
Therefore, it is more likely that the convergence problems of $M_{\rm inf}$ arise 
because it is this unknown parameter that determines the range of the distribution. 
The likelihoods associated to such probability distributions are known as 
{\it non-regular} (see, e.g., Smith 1985) and both likelihood and
Bayesian estimators may be affected, requiring alternative techniques 
(see, e.g., Atkinson et al. 1991, Nadal \& Pericchi 1998).
whose discussion is outside the scopes of the present work.

In conclusion, the bayesian estimators considered here cannot constrain 
the value of the $M_{\rm inf}$ parameter, and it also appears that our data are not 
yet strong enough to override the prior of the powerlaw slope. Therefore,
the value of $\alpha$ is sensitive to the choice of priors and their range 
(mostly on $M_{\rm inf}$), and the present data do not allow us to draw 
statistically robust conclusions on possible differences between the two 
SDP fields, if we allow $M_{\rm inf}$ to be a free parameter. In fact, the uncertainty on 
the estimated value of $\alpha$ should be that derived from scatter plots
like the one shown in Fig.~\ref{fig:comparisonl30}, rather than the formal
errors estimated by a specific method. 

%
%
\begin{figure}
%
%
%
%
%
%
%
%
\centering
%
\includegraphics[width=8.0cm,angle=0]{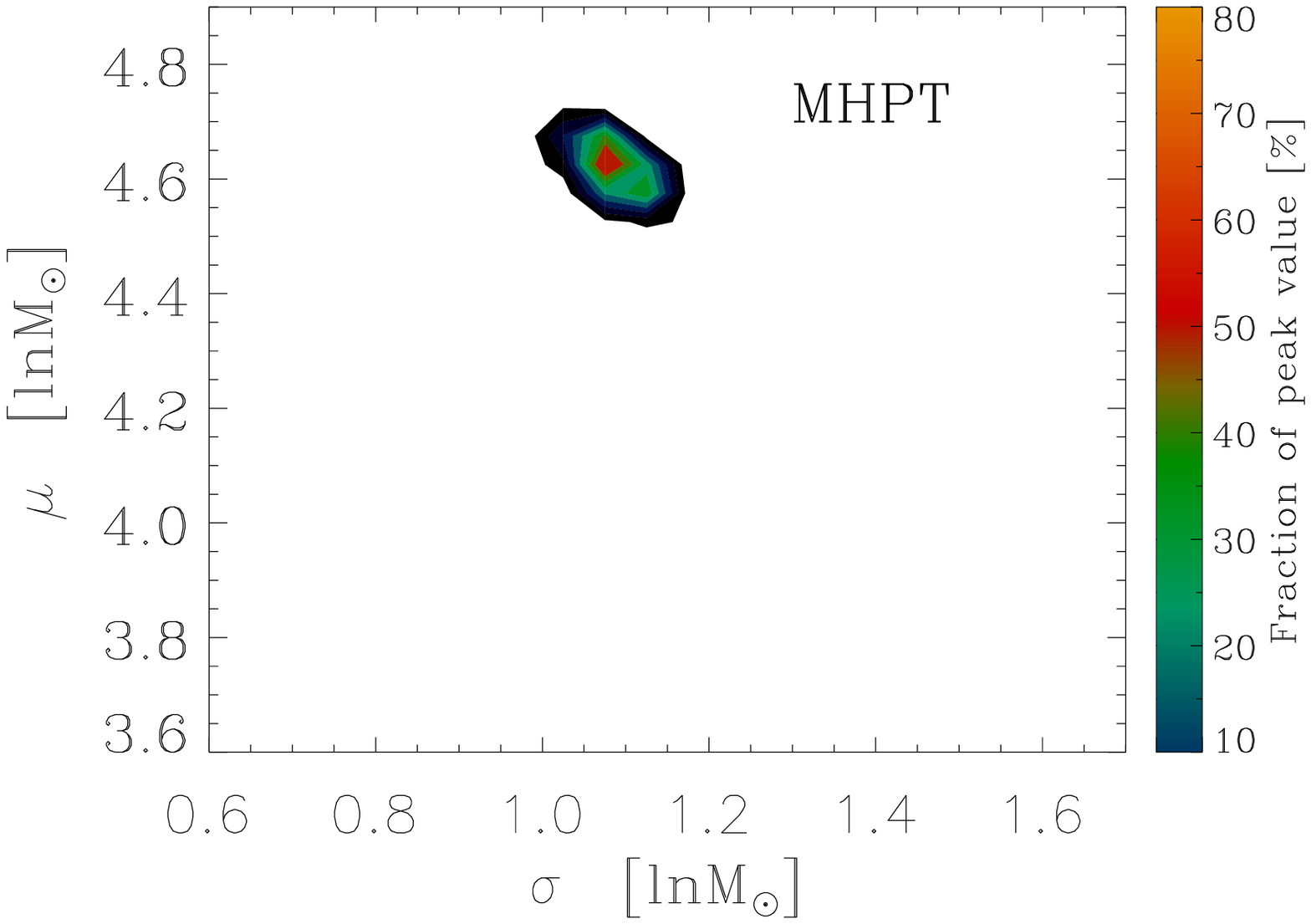}
\includegraphics[width=8.0cm,angle=0]{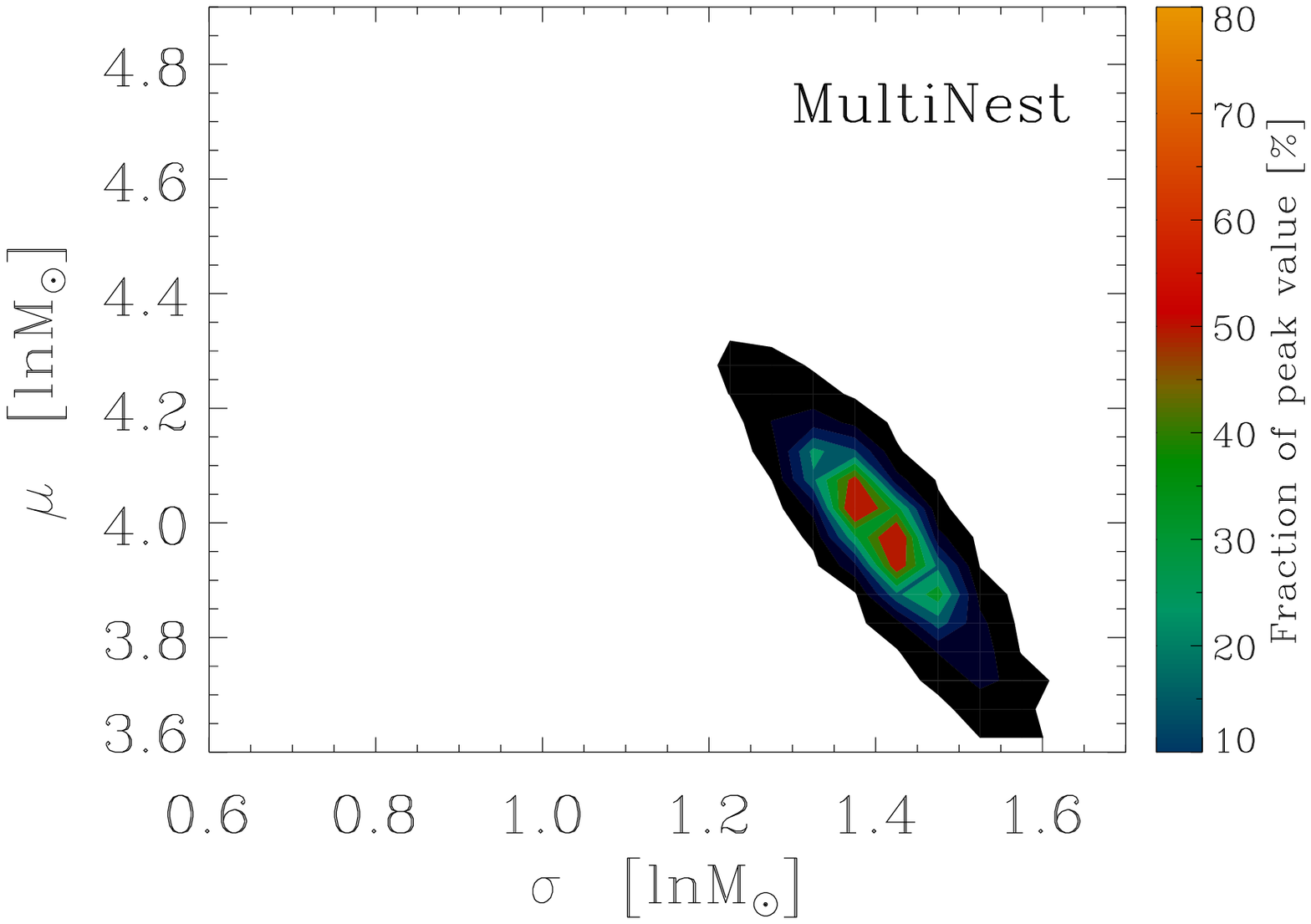}
\includegraphics[width=8.0cm,angle=0]{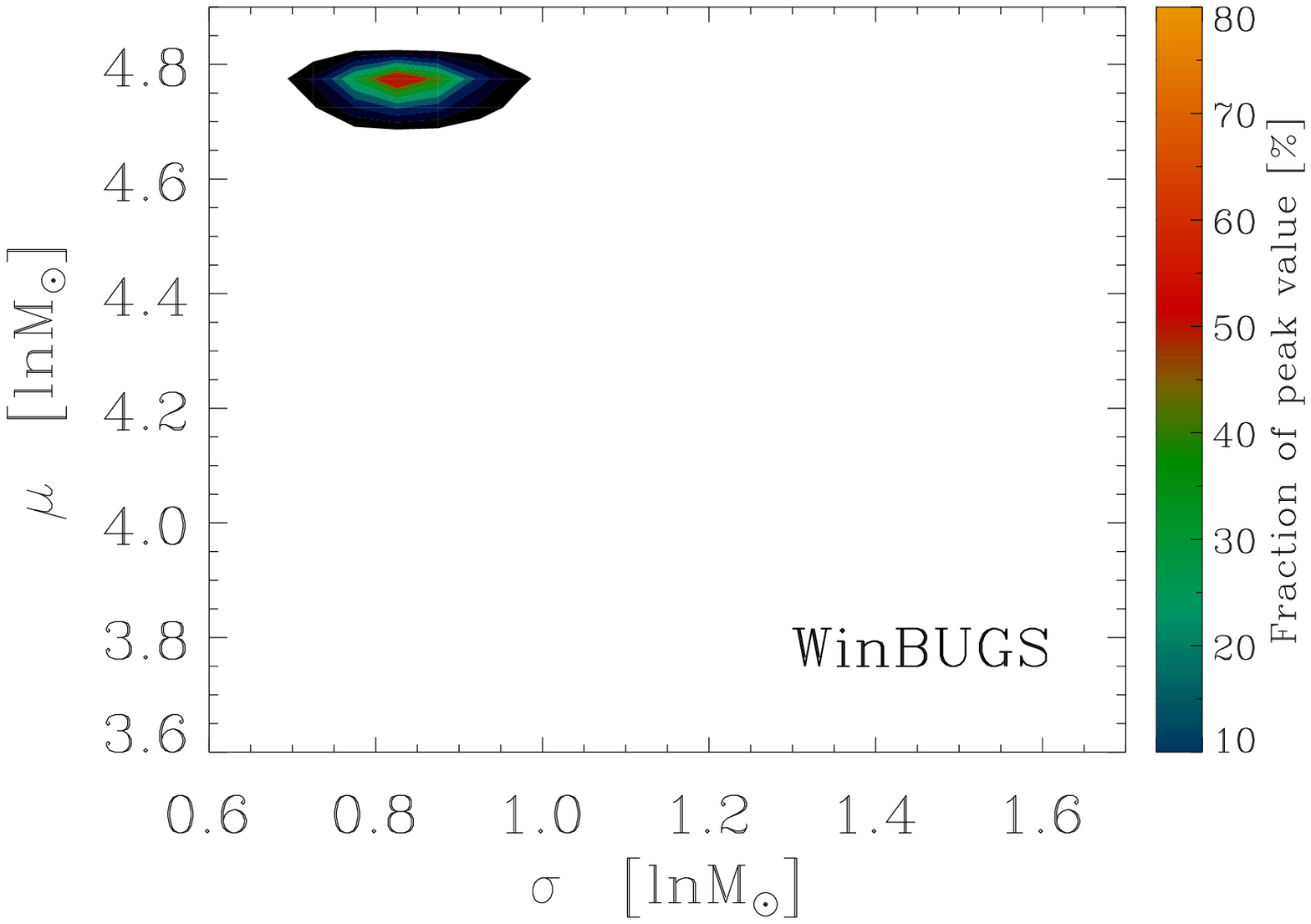}
%
\caption[ ]{
Results on the posterior distributions of the $\mu$ and $\sigma$
parameters for the lognormal PDF, shown simultaneously in
the $\mu-\sigma$ plane, for the MHPT (top panel),
MultiNest (middle panel) and WinBUGS (bottom panel) methods.
The results refer to the $\ell=30^{\circ}$ field.
}
\label{fig:l30log}
\end{figure}

%
%
\begin{figure}
%
%
\centering
%
\includegraphics[width=8.0cm,angle=0]{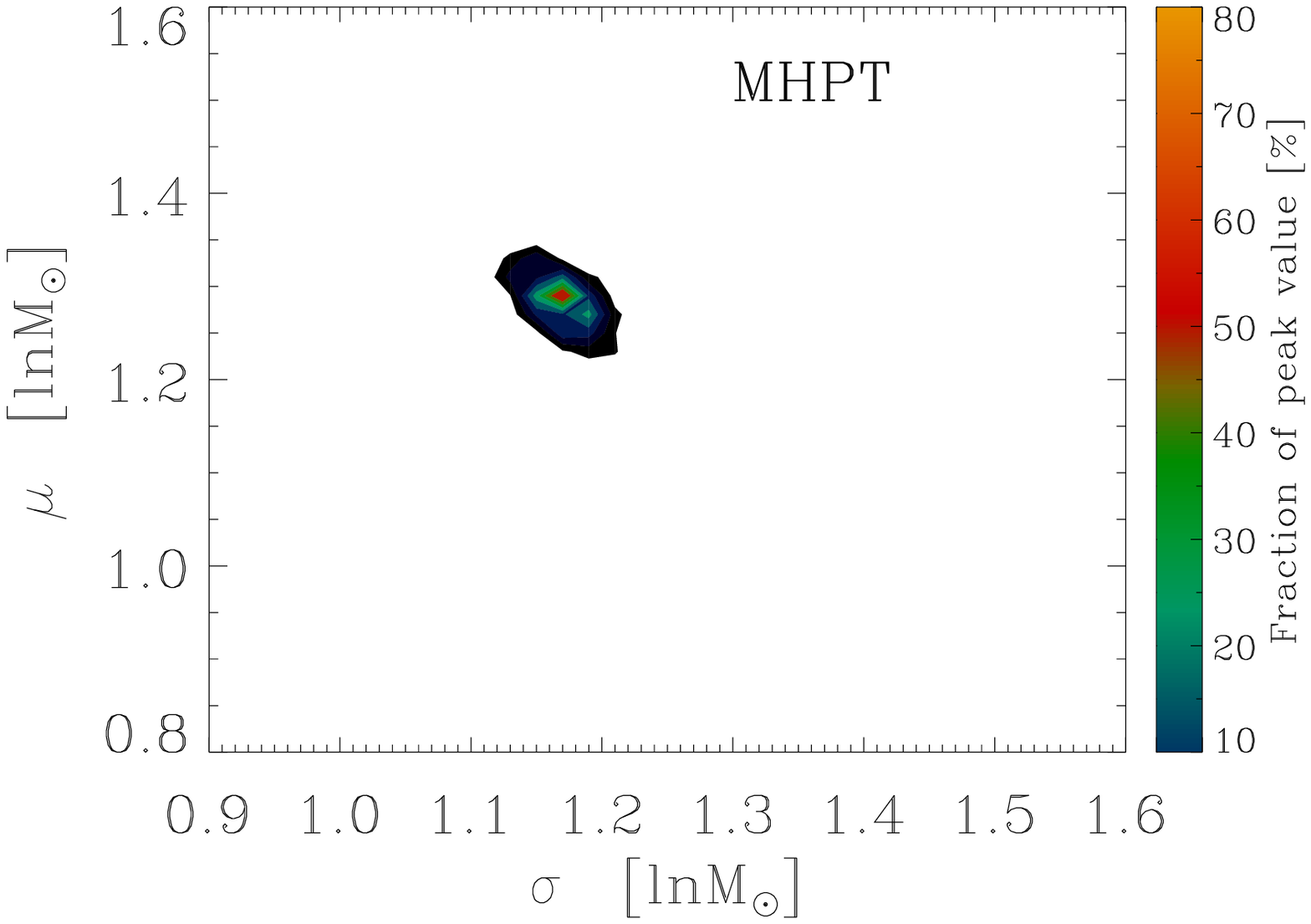}
\includegraphics[width=8.0cm,angle=0]{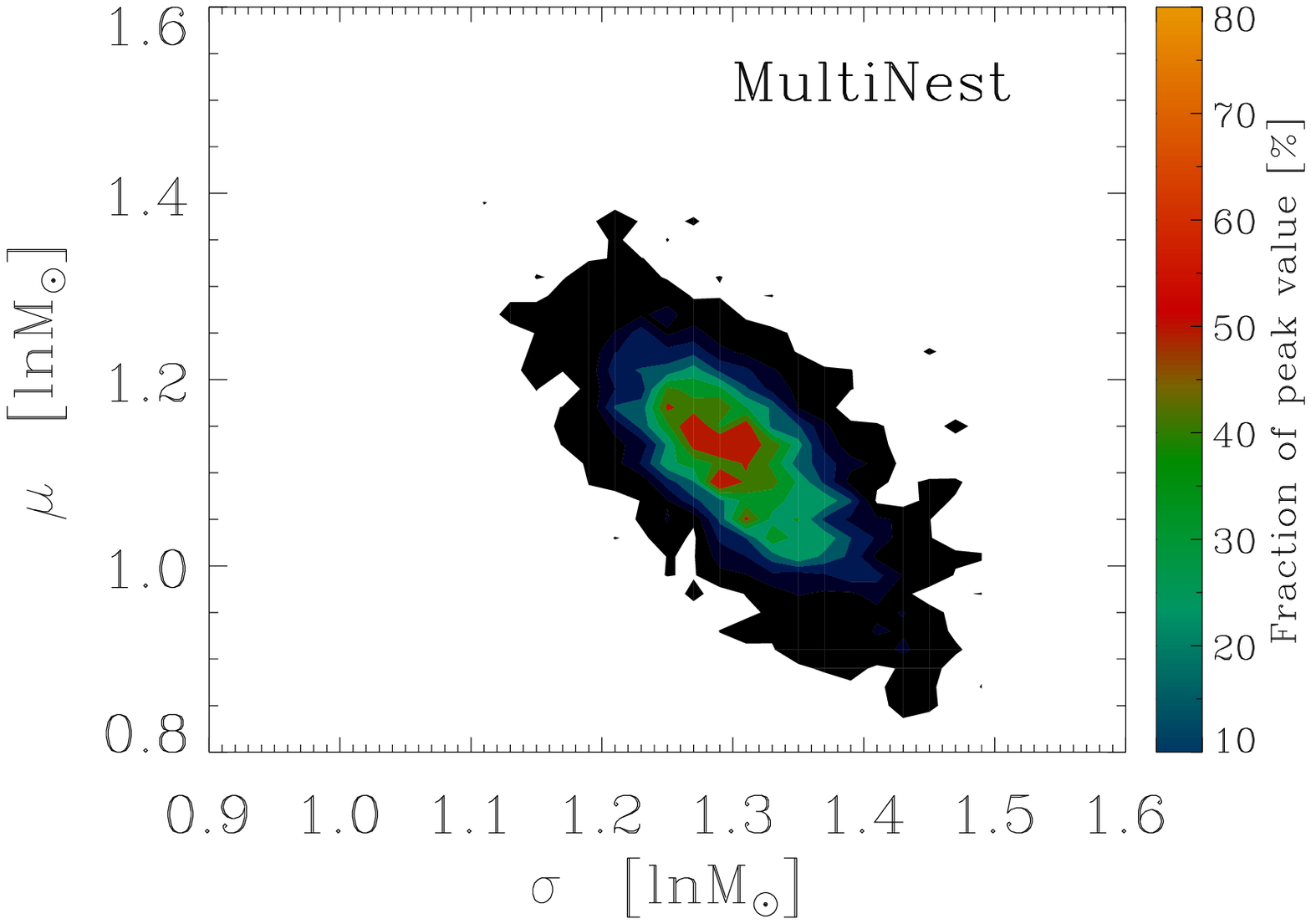}
\includegraphics[width=8.0cm,angle=0]{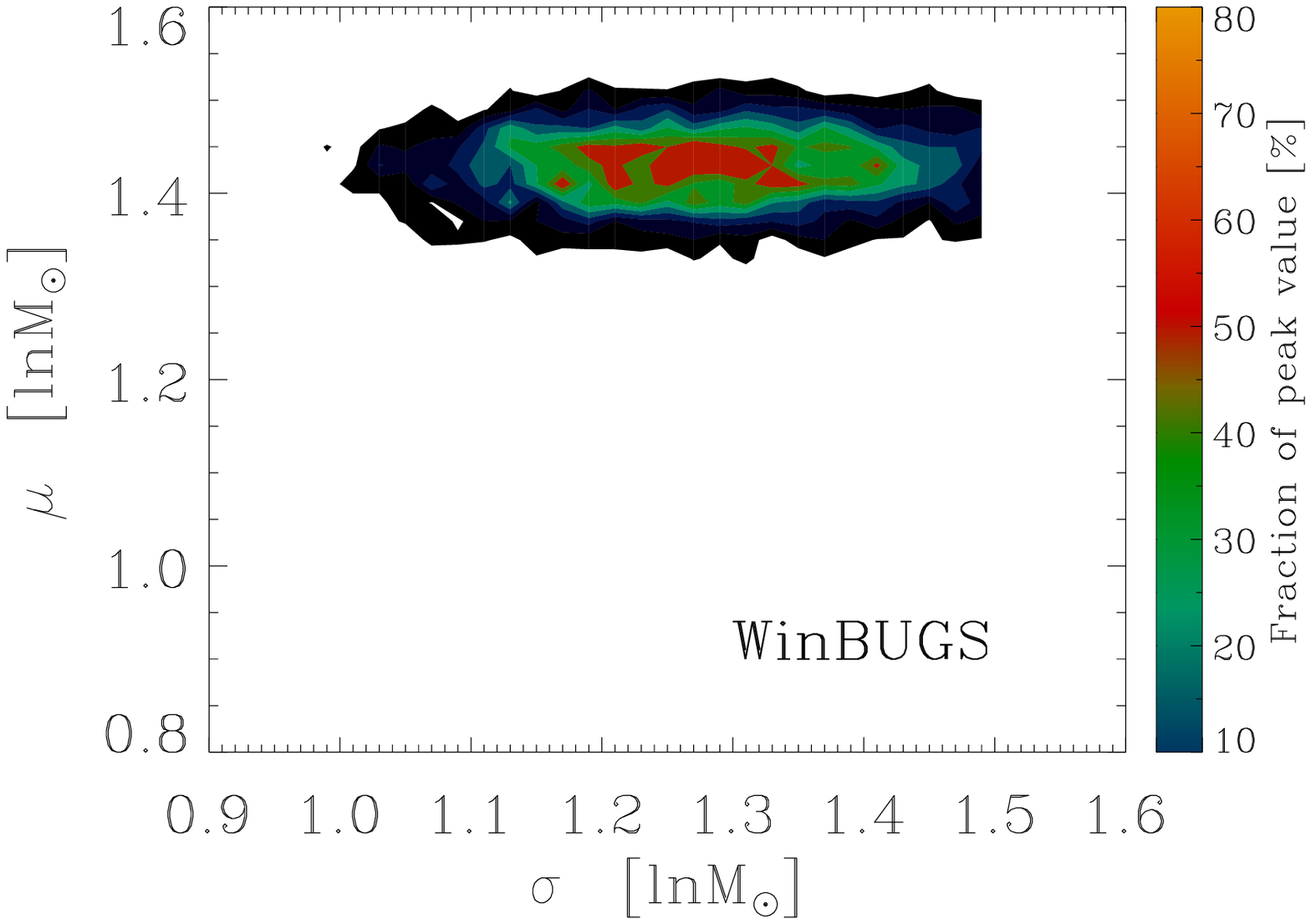}
%
\caption[ ]{
Same as Fig.~\ref{fig:l30log} for the $\ell=59^{\circ}$ field.
}
\label{fig:l59log}
\end{figure}

\subsubsection{Regular likelihood: keeping the value of $M_{\rm inf}$ fixed}
\label{sec:minfix}

In order to remove the potential problems associated with non-regular likelihoods, we carried out
some tests to determine whether the sensitivity to the prior range would be the same 
even when the value of the $M_{\rm inf}$ parameter is kept fixed.
We therefore modified the MHPT, MultiNest and WinBUGS procedures to have only one free
parameter, i.e., $\alpha$ in the case of the powerlaw model. $M_{\rm inf}$ was kept fixed,
but the range of the uniform prior on $\alpha$ was allowed to vary. 

In Fig.~\ref{fig:comparisonl30Minfix} we show the results obtained for the 
$\ell=30^{\circ}$ field. We selected a series of values for $M_{\rm inf}$, and then for each
of these values we run the three methods discussed above, repeating each procedure
several times using a different range $[\alpha_1,\alpha_2]$ for the uniform prior 
on $\alpha$. The figure shows three main features: {\it (i)} the two MCMC methods
(MHPT and WinBUGS) yield almost identical results for $\alpha$ 
(in Fig.~\ref{fig:comparisonl30Minfix} their corresponding symbols almost exactly overlap), 
independently of the selected value of $M_{\rm inf}$, whereas MultiNest progressively diverges 
from the other two methods. Then, {\it (ii)} the estimated values of $\alpha$
become larger, for all methods, when $M_{\rm inf}$ is increased. Finally, {\it (iii)} 
for each specific value of $M_{\rm inf}$ all three methods are rather {\it insensitive} 
to variations of the prior range $[\alpha_1,\alpha_2]$ (in Fig.~\ref{fig:comparisonl30Minfix} 
the error bars representing the variations of $\alpha$ when using a different prior range 
are not shown because they are tipically contained within the symbol size).

Therefore, the sensitivity to the uniform priors range, that has been discussed in
Section~\ref{sec:minfree} and graphically shown in Fig.~\ref{fig:comparisonl30}, 
disappears when $M_{\rm inf}$ is fixed and the only free parameter left is $\alpha$.
This result would appear to confirm that the effects discussed in Section~\ref{sec:minfree}
are indeed a special consequence of the non-regularity of the likelihood.

\subsection{Lognormal results}
\label{sec:compLN}

\subsubsection{Non-regular likelihood: considering $M_{\rm inf}$ a free parameter}
\label{sec:lognminfree}

The results from the posterior distribution of the parameters are listed in 
Table~\ref{tab:logn}, and they
are also shown graphically in figures~\ref{fig:l30log} and \ref{fig:l59log}. 
Although for the lognormal case we do not show a scatter plot similar to  
Fig.~\ref{fig:comparisonl30}, we have equally noted a high sensitivity of all methods
to the range of the uniform priors on $M_{\rm inf}$. This must be taken into account
when comparing the results of Table~\ref{tab:logn} with those quoted in Paper I (Table~5). 
Similar to what happens for the powerlaw PDF (Section~\ref{sec:minfree}), even in the lognormal 
case all methods considered here are unable to constrain the $M_{\rm inf}$ parameter 
which makes the likelihood non-regular. 

Our comparison is thus limited to the $\mu$ and $\sigma$ parameters. Then, comparing
their values in Table~\ref{tab:logn} we note again that 
the three methods discussed in Section~\ref{sec:globlike} deliver similar values for the
$\mu$ and $\sigma$ parameters, except for $\sigma$ in the $\ell=30^{\circ}$ field, where 
the differences are somewhat larger (up to $\simeq 40$\% in the worst case).
This is also clearly visible in Fig.~\ref{fig:l30log}, where it can be seen that the 
two MCMC-based methods 
deliver somewhat higher values of the $\mu$ parameter, compared to MultiNest, and
WinBUGS yields a significantly lower value of $\sigma$. In both SDP fields, we also
note the different shape of the posteriors distribution, with the MHPT and MultiNest 
distribution having a similar shape, i.e., with comparable widths in $\mu$ and $\sigma$
(although with a different scale), while 
WinBUGS tends to have a flattened distribution along the $\sigma-$ axis.

Therefore, even for the lognormal case, if the parameter $M_{\rm inf}$ is allowed 
to vary  the data are not constraining enough to allow one to reliably predict the 
{\it absolute} values of some key observables discussed here.
However, our estimates are still good enough to allow a
relative comparison between the two SDP regions. In fact,
even accounting for the different posterior distributions obtained with
the various methods and the different choice of priors, the parameter $\mu$ results
substantially higher in the $\ell=30^{\circ}$ field than in the 
$\ell=59^{\circ}$ region. On the other hand, the values of the parameter $\sigma$
are much more alike between the two SDP fields. This result appears to confirm
our earlier conclusion from Paper I, i.e., the CMFs of the two SDP fields have 
very similar shapes but different mass scales which, according to the simulations
discussed in Paper I, cannot be explained by distance effects alone. 

\subsubsection{Regular likelihood: keeping the value of $M_{\rm inf}$ fixed}
\label{sec:lognminfix}

\begin{figure}
%
%
%
\centering
%
\includegraphics[width=9.2cm,angle=0]{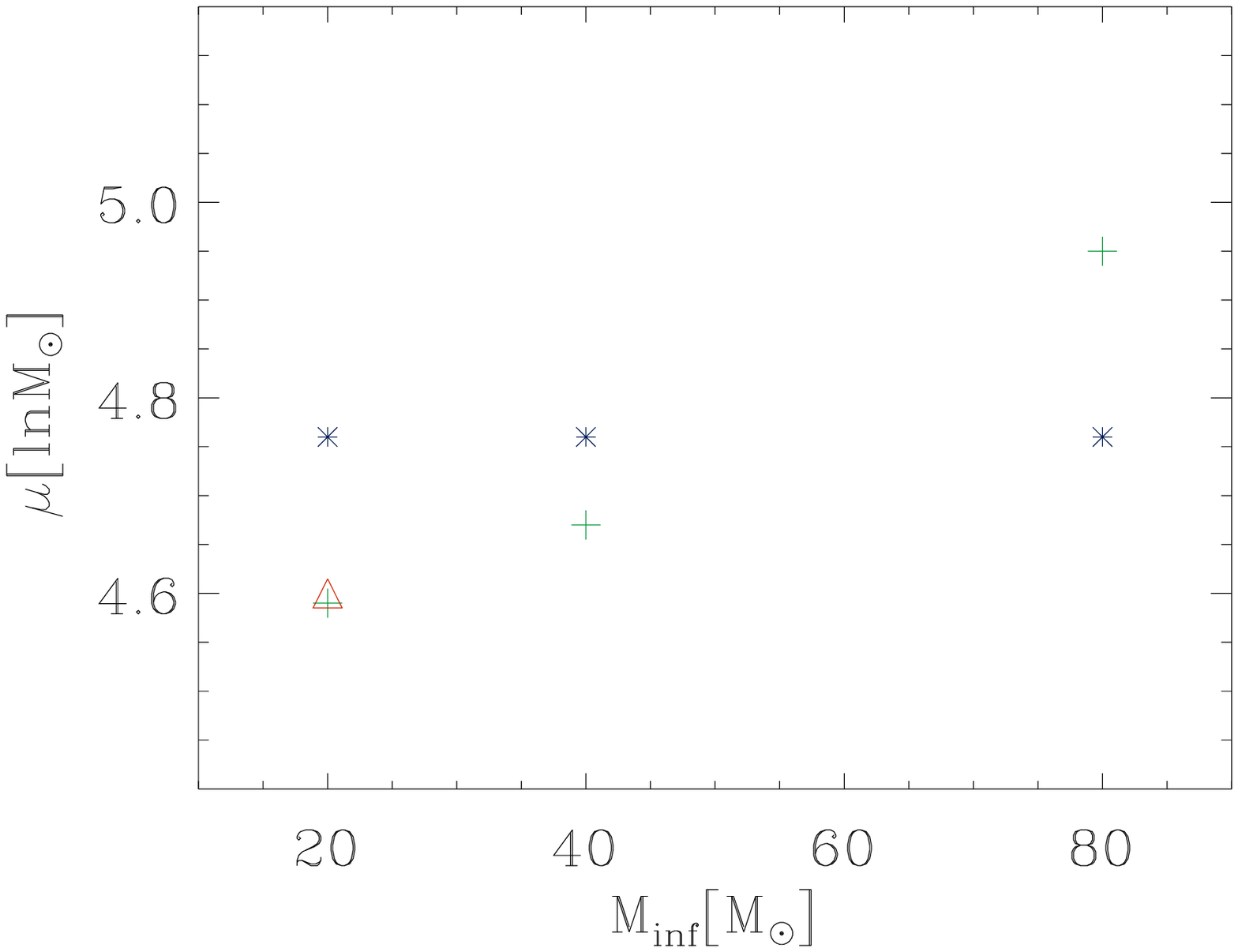}
\includegraphics[width=9.2cm,angle=0]{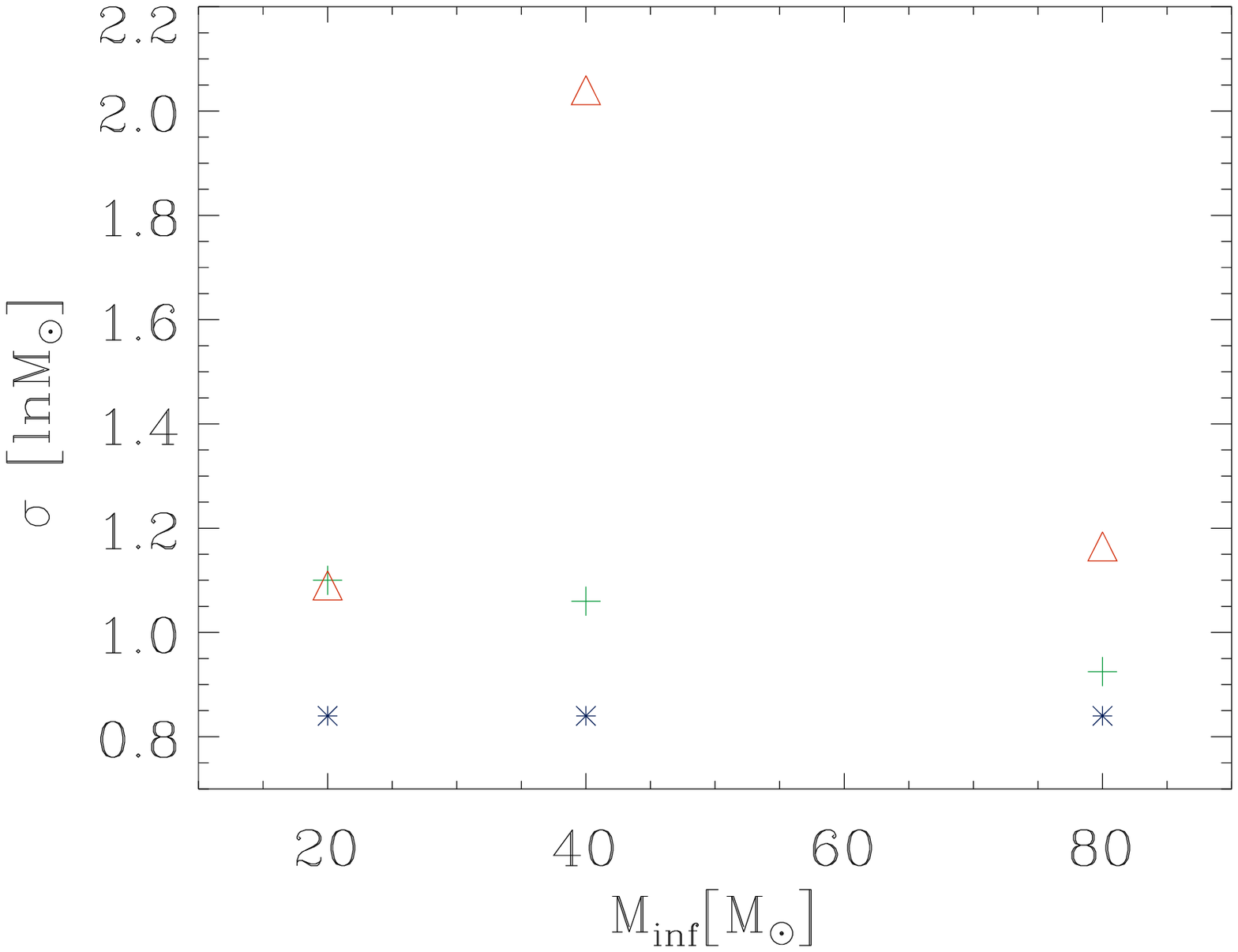}
%
\caption[ ]{
Comparison of results obtained for the parameters $\mu$ and $\sigma$ (lognormal case)
in the $\ell=30^{\circ}$ field keeping $M_{\rm inf}$ fixed. Symbols and colors are as
in Fig.~\ref{fig:comparisonl30}. 
}
\label{fig:logncomparisonl30Minfix}
\end{figure}

As already done in Section~\ref{sec:minfix} for the powerlaw model, we have run
similar tests also in the case of the lognormal distribution. Thus, we have selected 
some specific values of the $M_{\rm inf}$ parameter, and then for each
of these values we run the three methods discussed above, repeating each procedure
several times using different ranges $[\mu_1,\mu_2]$ and $[\sigma_1,\sigma_2]$ for the 
uniform priors on the $\mu$ and $\sigma$ parameters. 
The results for the $\ell=30^{\circ}$ field are shown in Fig.~\ref{fig:logncomparisonl30Minfix}.
As with the powerlaw case, the two MCMC methods yield similar values, although to a lesser
extent compared to Fig.~\ref{fig:comparisonl30Minfix}. The MultiNest algorithm converges
toward one end of the $[\mu_1,\mu_2]$ prior range when $M_{\rm inf} \ge 40\,M_\odot$, and this may 
also be the reason for the fluctuations seen in the parameter $\sigma$. We also note that
the parameter $\mu$ tends to increase with larger values of $M_{\rm inf}$, while $\sigma$ 
appears to be more stable, at least in the case of the MCMC methods. 

\begin{figure}
\centering
%
\includegraphics[width=8.0cm,angle=0]{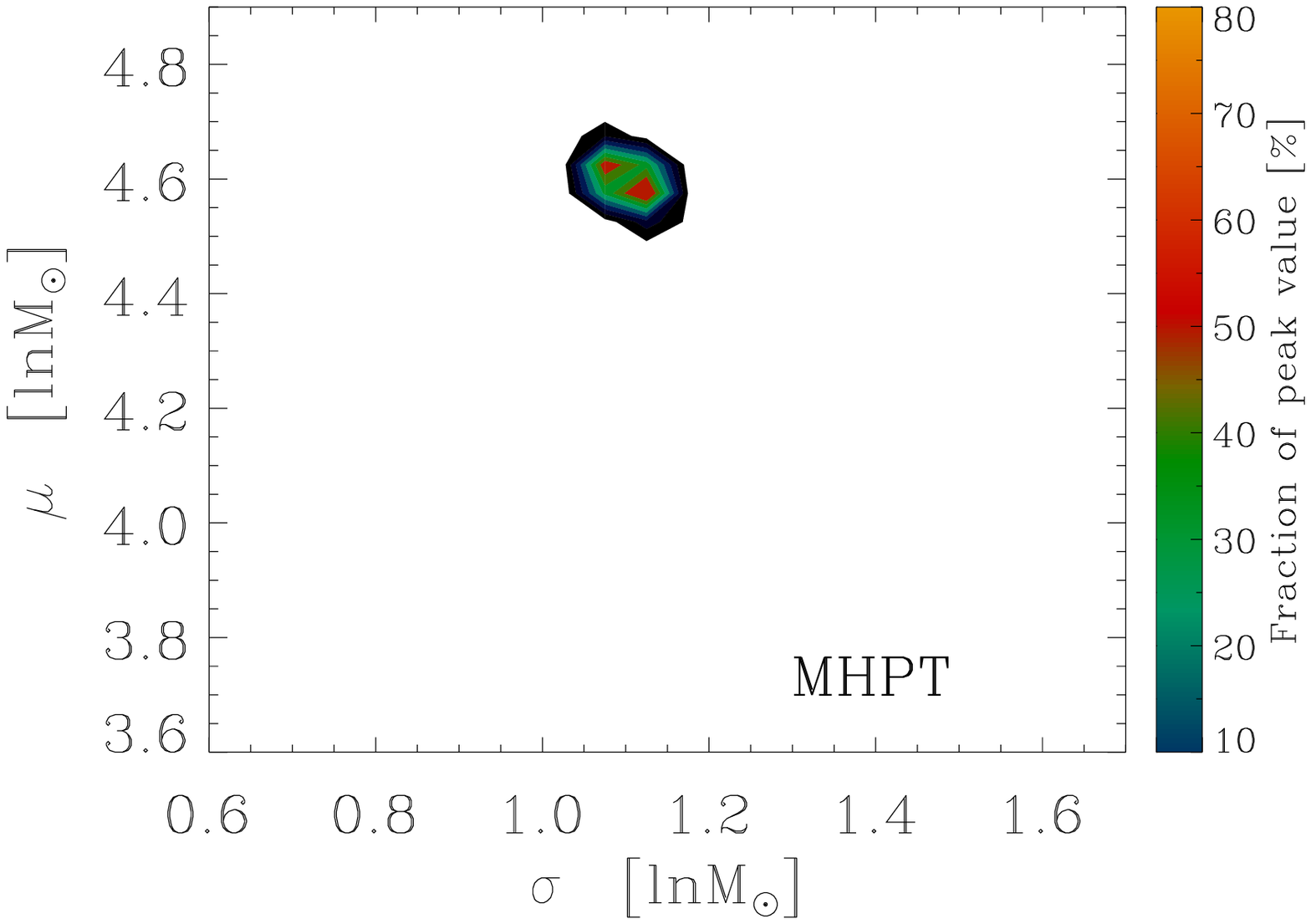}
\includegraphics[width=8.0cm,angle=0]{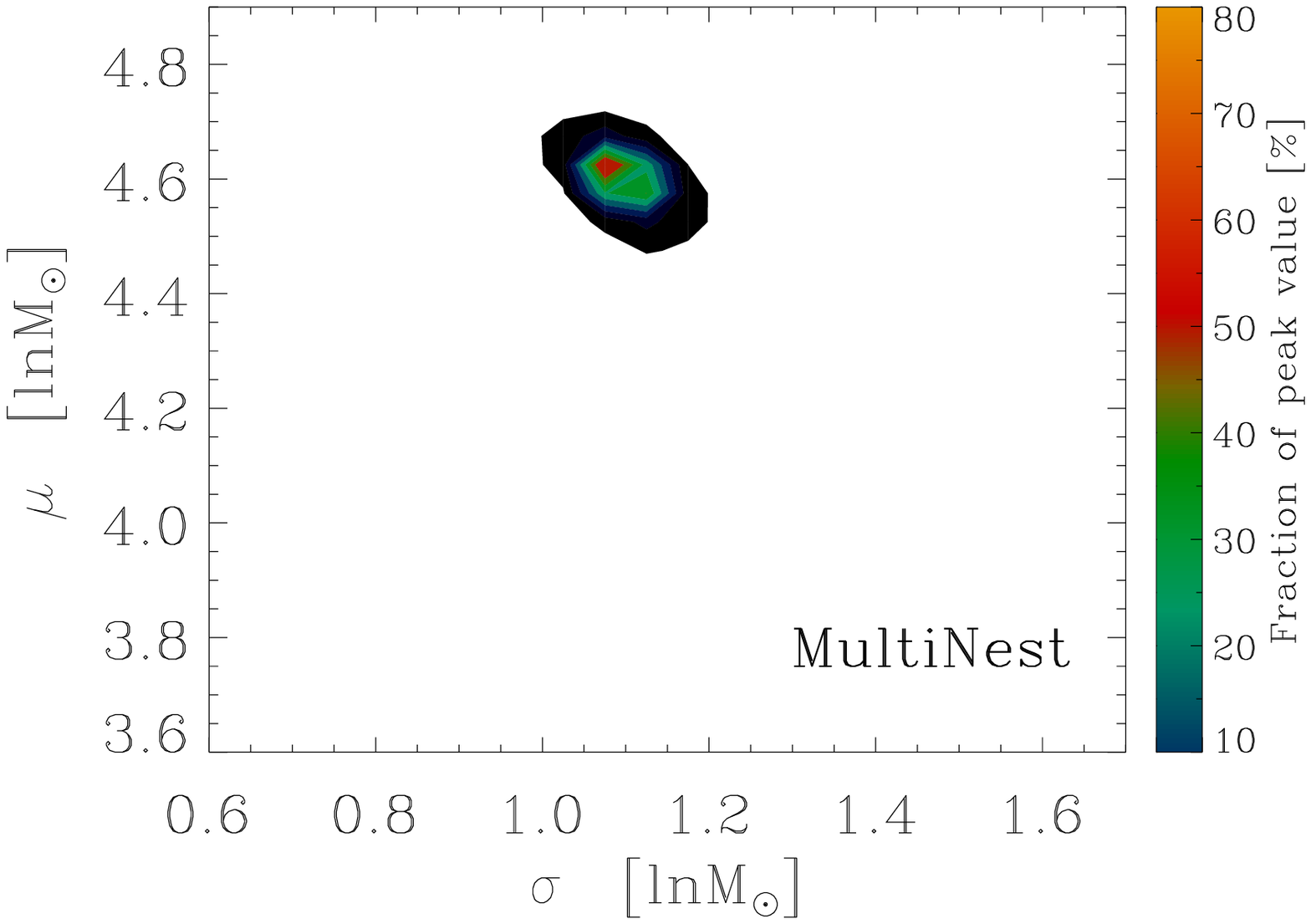}
\includegraphics[width=8.0cm,angle=0]{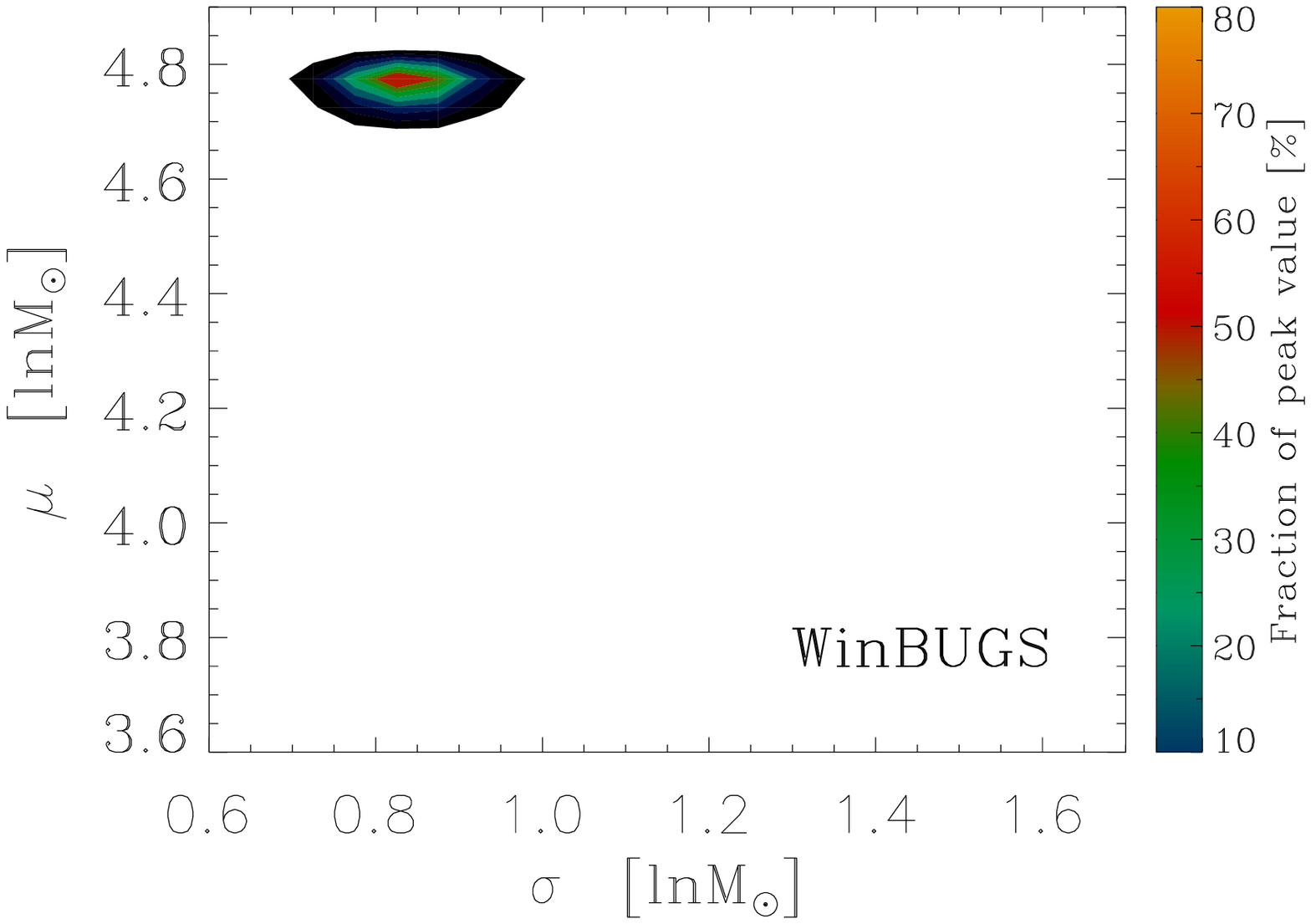}
%
\caption[ ]{
Same as Fig.~\ref{fig:l30log} for the case $M_{\rm inf}$ fixed ($M_{\rm inf}=20\,M_\odot$).
}
\label{fig:l30logMinfix}
\end{figure}

\begin{figure}
%
%
%
%
%
\centering
%
\includegraphics[width=8.0cm,angle=0]{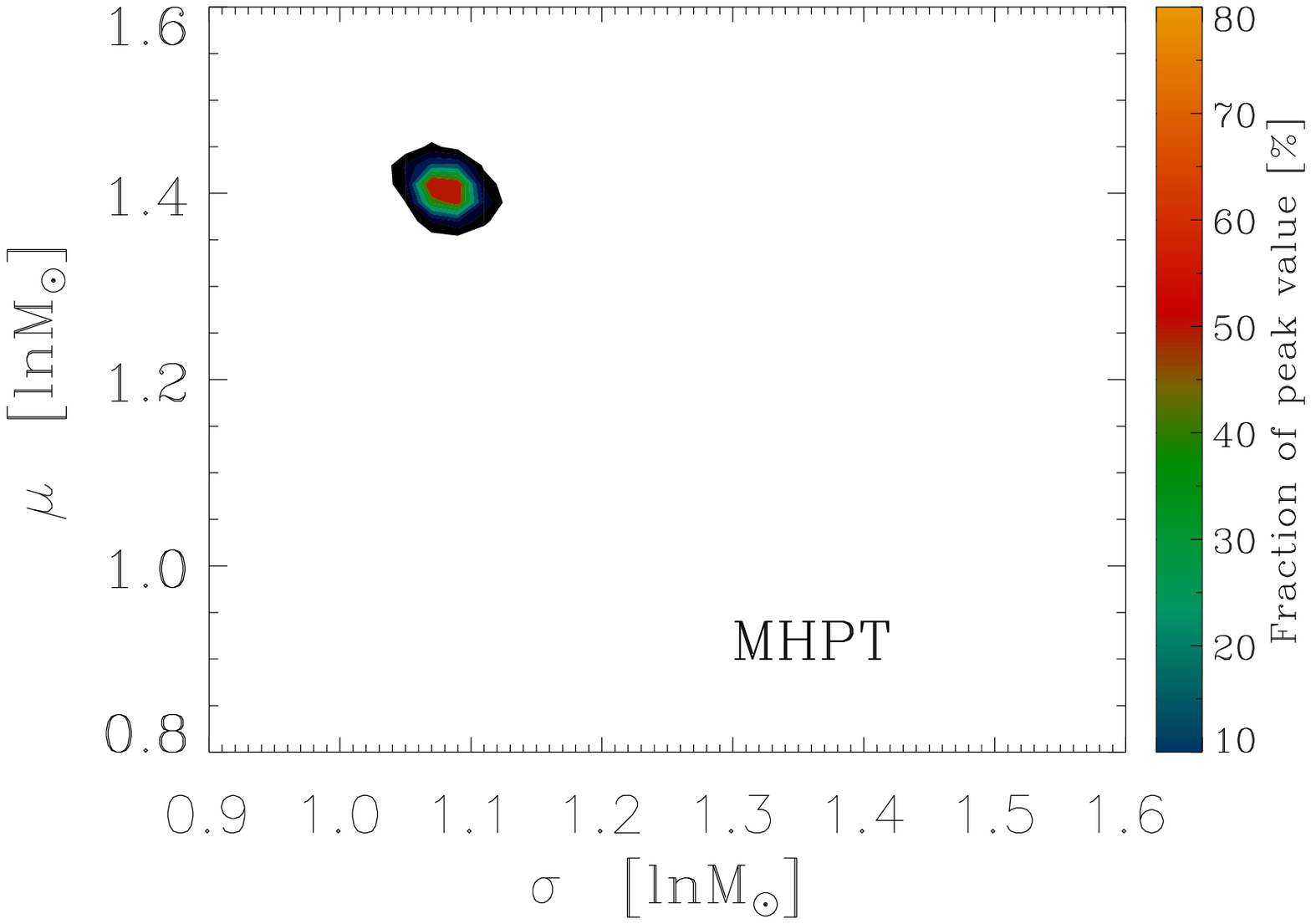}
\includegraphics[width=8.0cm,angle=0]{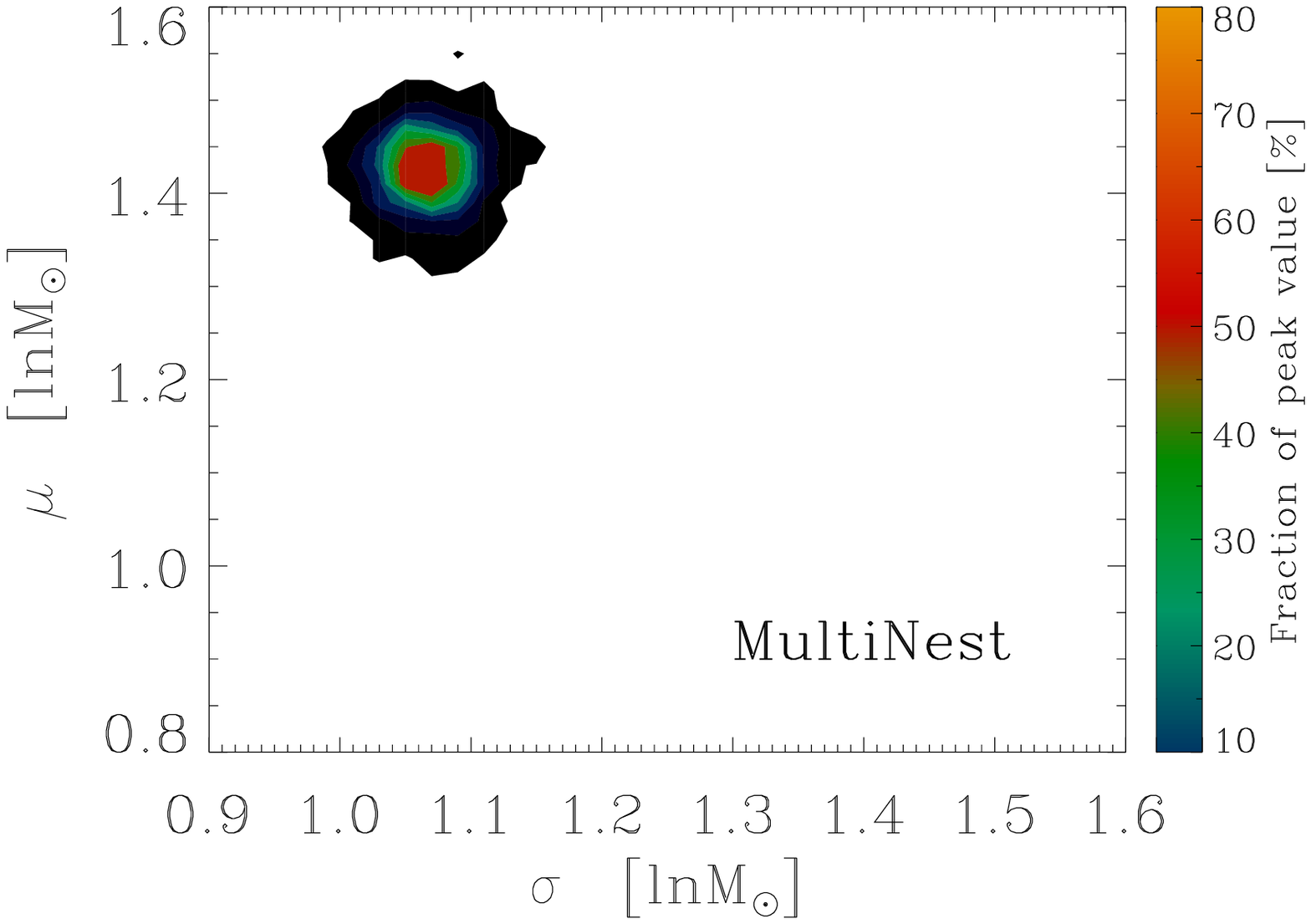}
\includegraphics[width=8.0cm,angle=0]{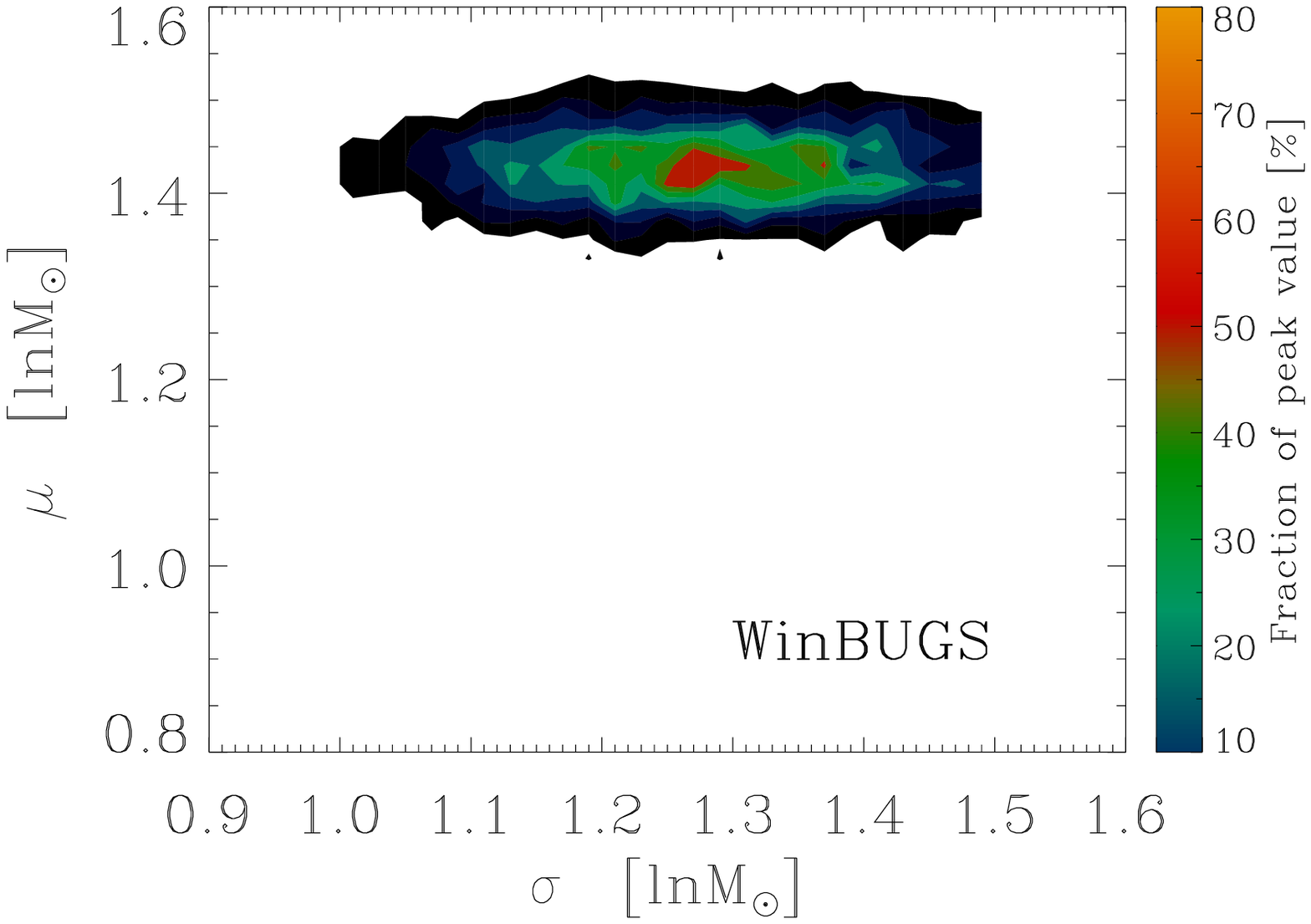}
%
\caption[ ]{
Same as Fig.~\ref{fig:l59log} for the case $M_{\rm inf}$ fixed ($M_{\rm inf}=0.5\,M_\odot$).
}
\label{fig:l59logMinfix}
\end{figure}

As with the powerlaw
model, when the parameter $M_{\rm inf}$ is fixed the posteriors of the remaining parameters, 
$\mu$ and $\sigma$, are {\it not} very sensitive to the ranges of the uniform priors, 
for all three methods.
It would thus appear that for the distributions analyzed here, the extreme sensitivity 
to the range of the uniform prior for the parameter $M_{\rm inf}$ is directly linked 
to the non-regularity of the likelihood, in the sense described in Section~\ref{sec:minfree}, 
rather than being related to the more general sensitivity of Bayesian inference to the 
choice of priors type and range.
As an additional comparison, in figures~\ref{fig:l30logMinfix} and \ref{fig:l59logMinfix}
we plot the posterior distributions in the $\mu-\sigma$ plane when $M_{\rm inf}$ is held 
fixed. 
Compared to figures~\ref{fig:l30log} and \ref{fig:l59log} one can note a better 
agreement among all algorithms and in particular between the MHPT and MultiNest methods,
while the WinBUGS distributions look very much the same.


%
%

%
%
%
%
\begin{table}
\caption{
Estimated Bayes factor, using the results listed in tables~\ref{tab:power}
and \ref{tab:logn}, for the case where $M_{\rm inf}$ is a free parameter. Positive
values of $\ln(BF)_{\rm ln/pw}$ favour the lognormal model over the powerlaw one.
}
\label{tab:globlike}
\centering
\begin{tabular}{lccr}
\hline\hline
               & \multicolumn{1}{c}{\bf $\ell=30^{\circ}$ field}  &
               & \multicolumn{1}{c}{\bf $\ell=59^{\circ}$ field}  \\
Method             & $\ln(BF)_{\rm ln/pw}$         &   & $\ln(BF)_{\rm ln/pw}$  \\
\hline
%
HME                & 2584           &     & 1381  \\
Laplace-Metropolis & 324            &     & 159   \\
MHPT               & 143            &     & 363   \\
MultiNest          & 114            &     & 109   \\
\hline
\end{tabular}
\end{table}

%
%
%
%
%
%
\begin{table}
\caption{
Estimated Bayes factor for the case where $M_{\rm inf}$ is fixed ($M_{\rm inf} = 20\,M_\odot$
and $M_{\rm inf} = 0.5\,M_\odot$, for the $\ell=30^{\circ}$ and $\ell=59^{\circ}$  fields, respectively; see
figures~\ref{fig:comparisonl30Minfix} and \ref{fig:logncomparisonl30Minfix}).
}
\label{tab:globlikeMinfFIXED}
\centering
\begin{tabular}{lccr}
\hline\hline
               & \multicolumn{1}{c}{\bf $\ell=30^{\circ}$ field}  &
               & \multicolumn{1}{c}{\bf $\ell=59^{\circ}$ field}  \\
Method             & $\ln(BF)_{\rm ln/pw}$         &   & $\ln(BF)_{\rm ln/pw}$  \\
\hline
%
HME                & 2675           &     & 1418 \\       
Laplace-Metropolis & $-$\tablefootmark{a}            &     & 145  \\
MHPT               & 189            &     & 586   \\      
MultiNest          & 356            &     & 171  \\      
%
\hline
\end{tabular}
\tablefoot{
\tablefoottext{a}{
No convergence obtained.}
}
\end{table}

%
%

%
%

\subsection{Model comparison}
\label{sec:compmod}

We have previously seen how the posterior distributions depend on the type 
of the parameters priors used and on their range, besides to depend on the specific
algorithm and software used to estimate the posteriors. 
As we mentioned in Section~\ref{sec:bayesintro}, the situation is even more
critical in model comparison, where the global
likelihood, a priors-weighted average likelihood,  and the Bayes factor will also 
depend on the choice of priors. In fact, even more so given that we usually
wish to compare models with different number of parameters.

Therefore, the results listed in Table~\ref{tab:globlike} should be regarded
with some caution. In the table we show the resulting Bayes factors (in logarithmic units) 
estimated using the global likelihoods reported in tables~\ref{tab:power} 
and \ref{tab:logn}. According to our discussion in Section~\ref{sec:ModComp}, 
the Bayes factors estimated by our analysis support the lognormal vs. the
powerlaw model. In Table~\ref{tab:globlike} we can note several features. 
{\it (i)} All methods result in the same conclusion, for both SDP regions.
{\it (ii)} However, the values of $\ln(BF)_{\rm ln/pw}$ are quite large: in the so-called
``Jeffreys scale'' a value of $\ln(BF)_{\rm 21}>5$ should be interpreted as
``strong'' support in favour of model 2 over model 1. The values that we find
are suspiciously large and suggest that they may be a consequence of the choice
of priors. 
{\it (iii)} With the exception of MultiNest, all other method show a substantial 
difference in the Bayes factors estimated for the two SDP fields. 

We have estimated the Bayes factors also for the case of the parameter
$M_{\rm inf}$ fixed, in order to check for any significant difference. 
The results are shown in Table~\ref{tab:globlikeMinfFIXED}, and we can note that
the values of $\ln(BF)_{\rm ln/pw}$ are still positive and still quite large.
The values $\ln(BF)_{\rm ln/pw}$ are also strongly dependent on the selected
value of the $M_{\rm inf}$ parameter, and are typically lower for higher values
of $M_{\rm inf}$.
Therefore, our preliminary conclusion is that the {\it lognormal} model appears to
better describe the CMF measured in the two SDP regions. However, we caution
that this conclusion may be affected by different choices of priors and their ranges.

\section{Conclusions}
\label{sec:conclusions}

Following our study (Paper I) of the two Hi-GAL, SDP fields centered at $\ell=59^{\circ}$ 
and $\ell=30^{\circ}$, we have applied a full Bayesian analysis to the 
CMF of these two regions to determine how well two simple models, powerlaw 
and lognormal, describe the data. First, we have determined the Bayesian posterior 
probability distribution of the model parameters. Next, we have carried out a 
quantitative comparison of these models, given data and an explicit set of assumptions.
This analysis has highlighted the peculiarities of Bayesian inference compared to
more commonly used MLE methods. In parameters estimation, Bayesian inference allows 
to estimate the probability distributions of each parameter, making it easier in principle
to obtain realistic error bars on the results and, in addition, to include prior
information on the parameters. However, Bayesian inference may be computationally intensive,
and we have also shown that the results may be quite sensitive to the priors type and range, 
particularly if the parameters limit the range of the distribution (such as the 
$M_{\rm inf}$ parameter).

In terms of the powerlaw model,
we have found that the three bayesian methods described here deliver remarkably similar 
values of the powerlaw slope, for both SDP fields. Likewise,
for the lognormal model of the CMF, we have found that the three Bayesian methods
deliver similar values for the $\mu$ (center of the lognormal distribution) and $\sigma$ 
(width of the lognormal distribution) parameters, separately for both SDP fields.
In addition, the parameter $\mu$ results substantially higher in the $\ell=30^{\circ}$
field than in the $\ell=59^{\circ}$ region, while the values of the parameter $\sigma$
are much more alike between the two SDP fields. This result confirms
our earlier conclusion from Paper I, i.e., the CMFs of the two SDP fields have
very similar shapes but different mass scales.
We have also shown that the difference with respect to the values of the parameters
determined in Paper I may be due to the sensitivity of the
posterior distributions to the specific choice of the parameters priors, and in
particular of the $M_{\rm inf}$ parameter.

As far as model comparison is concerned, we have discussed and compared several methods to
compute the global likelihood, which in general cannot be calculated analytically
and is fundamental to estimate the Bayes factor. All methods tested here showed 
that the lognormal model appears to better describe the CMF measured in the 
two SDP regions. However, this preliminary conclusion is dependent 
on the choice of parameters priors and needs to be confirmed using more constraining data. 

\appendix

\section{Procedure to implement the Metropolis-Hastings algorithm with parallel tempering (MHPT) }
\label{sec:procMHPT}

We briefly list here the  main steps of the Metropolis-Hastings algorithm with the inclusion of parallel tempering
(see sections~\ref{sec:defs} and \ref{sec:MH} for defintions).  

\vspace*{2mm}
\noindent
1:   Initialize the parameters vector, for all tempered distributions \\
     \hspace*{3mm}   $ \mbox{{\boldmath $\theta$}$_{0,i}$} = \mbox{{\boldmath $\theta$}$_0$} $, \hspace*{1mm} $1 \le i \le n_\beta$ \\
2:   Start MCMC loop \\
     \hspace*{3mm}   {\bf for $t=1,...,(T-1)$ } \\
3:   \hspace*{4mm}   Start {\it parallel tempering} loop \\
     \hspace*{7mm}   {\bf for $i=0,1,...,(n_\beta-1)$ } \\
4:   \hspace*{8mm}  Propose a new sample drawn from a Normal \\ 
     \hspace*{11mm}  distribution with mean equal to current \\
     \hspace*{11mm}  parameters values and standard deviation fixed \\
     \hspace*{12mm}  $\mbox{{\boldmath $\theta$}$_{\it prop}$} \sim N(\mbox{{\boldmath $\theta$}$_{t,i}$}; \mbox{{\boldmath $\sigma$}}) $ \\
5:   \hspace*{8mm}  Compute the {\it Metropolis ratio} using Eq.~(\ref{eq:posteriorbeta}) \\
     \hspace*{12mm}  $\ln r = \ln P(\mbox{{\boldmath $\theta$}$_{\it prop}$} | D, {\mathcal M}, \beta_i) -
                     \ln P(\mbox{{\boldmath $\theta$}$_{t,i}$} | D, {\mathcal M}, \beta_i) $ \\
6:   \hspace*{8mm}  Sample a uniform random variable \\
     \hspace*{12mm}  $ u_1 \sim Uniform(0,1)$ \\
7:   \hspace*{8mm}  {\bf if $\ln u_1 \le \ln r$ then } \\
     \hspace*{14mm}  $\mbox{{\boldmath $\theta$}$_{t+1,i}$} = \mbox{{\boldmath $\theta$}$_{\it prop}$} $ \\
     \hspace*{12mm}  {\bf else } \\
     \hspace*{14mm}  $\mbox{{\boldmath $\theta$}$_{t+1,i}$} = \mbox{{\boldmath $\theta$}$_{t,i}$} $ \\
     \hspace*{12mm}  {\bf end if} \\
8:   \hspace*{5mm}   {\bf end for}  \hspace*{3mm} End parallel tempering loop \\
9:   \hspace*{5mm}  Sample another uniform random variable \\
     \hspace*{10mm}  $ u_2 \sim Uniform(0,1)$ \\
10:  \hspace*{3mm}  Do swap between chains? \\
     \hspace*{10mm}  ($n_{\rm swap}$ = N. of swaps between chains) \\
11:  \hspace*{4mm}  {\bf if $u_2 \le 1/n_{\rm swap}$ then } \\
12:  \hspace*{7mm}  Select random chain: \\
     \hspace*{12mm}  $j \sim UniformInt(1,n_\beta-1) $ \\
13:  \hspace*{7mm}  Compute $r_{\rm swap}$ \\
     \hspace*{12mm}  $ \ln r_{\rm swap} = \ln P(\mbox{{\boldmath $\theta$}$_{t,j+1}$} | D, {\mathcal M}, \beta_j) + 
                       \ln P(\mbox{{\boldmath $\theta$}$_{t,j}$} | D, {\mathcal M}, \beta_{j+1}) $ \\
     \hspace*{12mm}  $ - \ln P(\mbox{{\boldmath $\theta$}$_{t,j}$} | D, {\mathcal M}, \beta_j) 
                       - \ln P(\mbox{{\boldmath $\theta$}$_{t,j+1}$} | D, {\mathcal M}, \beta_{j+1}) $ \\
14:  \hspace*{8mm}  $ u_3 \sim Uniform(0,1)$ \\
15:  \hspace*{8mm}  {\bf if $ \ln u_3 \le \ln r_{\rm swap}$ then } \\
     \hspace*{16mm} Swap parameters states of chains $j$ and $j+1$ \\
     \hspace*{16mm} $ \mbox{{\boldmath $\theta$}$_{t,j}$} \leftrightarrow \mbox{{\boldmath $\theta$}$_{t,j+1}$} $ \\ 
15:  \hspace*{8mm}  {\bf end if } \\
16:  \hspace*{4mm}  {\bf end if} \\
17:  \hspace*{2mm}  {\bf end for} \hspace*{3mm} End MCMC loop


\begin{acknowledgements}
L.O. would like to thank L. Pericchi for fruitful discussions on various issues
related to Bayesian inference and parameter priors.
\end{acknowledgements}

\bibliographystyle{aa} 
\bibliography{refs}    

\end{document}